**Crystal field excitations and magnons: their roles in oxyselenides $Pr_2O_2M_2OSe_2$ ($M$ = Mn, Fe)**

R. K. Oogarah[1], C. P. J. Stockdale[2], C. Stock[2], J. S. O. Evans[3], A. S. Wills[4], J. W. Taylor[5], E. E. McCabe*[1]

[1]School of Physical Sciences, Ingram Building, University of Kent, Canterbury, Kent, U.K.

[2]School of Physics and Astronomy, University of Edinburgh, Edinburgh, EH9 3JZ, U.K.

[3]Department of Chemistry, Durham University, Lower Mountjoy, South Road, Durham, DH1 3LE, U.K.

[4]Department of Chemistry, University College London, 20 Gordon Street, London, WC1H 0AJ, U.K.

[5]ISIS Facility, Rutherford Appleton Laboratory, Chilton, Didcot, OX11 0QX, U.K.

*e.e.mccabe@kent.ac.uk

We present the results of neutron scattering experiments to study the crystal and magnetic structures of the Mott-insulating transition metal oxyselenides $Pr_2O_2M_2OSe_2$ ($M$ = Mn, Fe). The structural role of the non-Kramers $Pr^{3+}$ ion is investigated and analysis of $Pr^{3+}$ crystal field excitations performed. Long-range order of $Pr^{3+}$ moments in $Pr_2O_2Fe_2OSe_2$ can be induced by an applied magnetic field.

## Introduction

Much research is directed towards understanding the parent phases of unconventional superconductors and in particular, their structures, exchange interactions and the driving forces for structural distortions. The phase diagram for cuprate superconductors derives from parent Mott-insulating antiferromagnetic (AFM) phases and unconventional superconductivity is induced by charge doping.[1-3] The iron-based superconducting systems are less clearly understood,[4-6] with poorly metallic or semimetallic parent phases, raising the question as to whether a localised or an itinerant/spin-density wave model is more appropriate.[7]

The "$M_2$O" oxyselenides e.g. $La_2O_2M_2OSe_2$ ($M$ = Mn, Fe, Co) adopt tetragonal crystal structures composed of alternating $[La_2O_2]^{2+}$ fluorite-like and $[M_2O]^{2+}$ layers separated by selenide anions (Figure 1a). This gives an unusual $M^{2+}$ coordination environment with pseudo-octahedral coordination by two oxide anions (in-plane) and four selenide anions (above and below the $[M_2O]^{2+}$ sheets), Figure 1b.[8] Electronic structure calculations on $La_2O_2Fe_2OSe_2$ reveal relatively narrow $Fe^{2+}$ 3d bands in this layered material, pointing to significant correlation effects.[9] This is consistent with electrical resistivity measurements[9-11] and suggests that these "$Fe_2O$" materials provide a suitable Mott-insulating reference system in which to study the exchange interactions and anisotropy in layered mixed-anion phases.[12]

The $La_2O_2M_2OSe_2$ ($M$ = Mn, Fe, Co) oxyselenides order antiferromagnetically (AFM) on cooling ($T_N$ = 160 – 168 K, 89 K and 220 K for Mn[13-15], Fe[16-17] and Co[18-20] analogues, respectively). There are three magnetic exchange interactions in the $M_2$O layers: nearest-neighbour (nn) AFM exchange $J_1$, next-nearest-neighbour (nnn) 180° $M$–O–$M$ AFM exchange $J_{2'}$ and nnn ~97° $M$–Se–$M$ exchange $J_2$ (Figure 1c). The relative strengths of these exchange interactions vary with $M$ electronegativity: for $M$ = Mn, nn $J_1$ interactions dominate, giving G-type AFM order with moments perpendicular to the $[M_2O]^{2+}$ layers (Figure 1d),[14-15] while for the more electronegative $M = Co^{2+}$ analogue, 180° $J_{2'}$ interactions dominate leading to a magnetic structure with $k$ vector $k$ = (½ ½ 0) with in-plane moments and nn moments perpendicular to one another.[18-19, 21-22] For $M$ = Fe, Fe–Se–Fe $J_2$ interactions are FM[10] (consistent with the FM chains of edge-linked $FeSe_4$ tetrahedra in $Ce_2O_2FeSe_2$).[23-24] $La_2O_2Fe_2OSe_2$ adopts a two-$k$-vector magnetic structure ($k$ = (½ 0 ½) and $k$ = (0½ ½)) with in-plane moments directed along the Fe–O bonds, with FM $J_2$ and AFM $J_{2'}$ interactions consistent with theory, and nearest-neighbour moments orthogonal to one another (Figure 1e) (referred to here as the 2$k$ structure).[11-12, 17, 25] Second-order exchange interactions alone are insufficient to stabilise this 2$k$ magnetic structure and higher-order terms (e.g. from $Fe^{2+}$ spin anisotropy) are needed. The onset of AFM order is 2D-Ising[12] like and inelastic neutron scattering (INS) experiments revealed an anisotropy gap (~5 meV at 2 K) and very weak magnetic exchange interactions,[17] consistent with band narrowing reported for these materials.[9]

Lanthanide magnetism can also influence the magnetism of the transition metal sublattice, with $Fe^{2+}$ spin reorientation driven by lanthanide magnetic order in 1111 materials PrFeAsO and NdFeAsO.[26-27] In the layered copper oxide superconducting systems, the electronic state of $Pr^{3+}$ ions (leading to magnetic or non-magnetic ground states) was key to whether copper-based superconductivity evolved on cooling. [28-30] This has prompted significant research into possible coupling between the lanthanide and transition metal sublattices in layered systems.[24, 31-45] Probing the crystal field excitations (CEFs) of the lanthanide ions by inelastic neutron scattering (INS) experiments has given insights into both the lanthanide magnetism and the transition metal sublattice.[41, 46] In addition, the non-Kramers $4f^2$ ($^3H_4$) $Pr^{3+}$ ion is also susceptible to symmetry-lowering structural distortions driven by 4f electron degrees of freedom.[32] In PrMnAsO and PrMnSbO, structural distortions occur at ~35 K which lower the symmetry of the $Pr^{3+}$ site from 4*mm* to *mm*2, accompanied by long range order of $Pr^{3+}$ moments and reorientation of $Mn^{2+}$ moments.[32-33] An analogous structural distortion occurs in $Pr_2O_2Mn_2OSe_2$ below 36 K, lowering the crystal symmetry from *I*4/*mmm* to *Immm*, but, unlike PrMnAsO and PrMnSbO, there is no evidence for long-range ordering of $Pr^{3+}$ moments in $Pr_2O_2Mn_2OSe_2$.

In this work we investigate the structural changes in $Pr_2O_2Fe_2OSe_2$ on cooling and observe a subtle orthorhombic structural distortion at ~23 K. We compare the low temperature NPD data collected in zero- and 5 T applied fields to study the metamagnetic phase transition in this material. Using INS, we investigate the magnetism in $Pr_2O_2Mn_2OSe_2$ and confirm the lack of an anisotropy gap for this $Mn^{2+}$ material, in contrast to $Fe^{2+}$ analogues.[17] We also investigate the local electronic crystalline electric field environment by studying the low-energy CEFs of the $Pr^{3+}$ ion for both *M* = Mn, Fe materials. We show how the transition metal magnetic excitations might couple with the $Pr^{3+}$ CEFs and influence the structural behaviour.

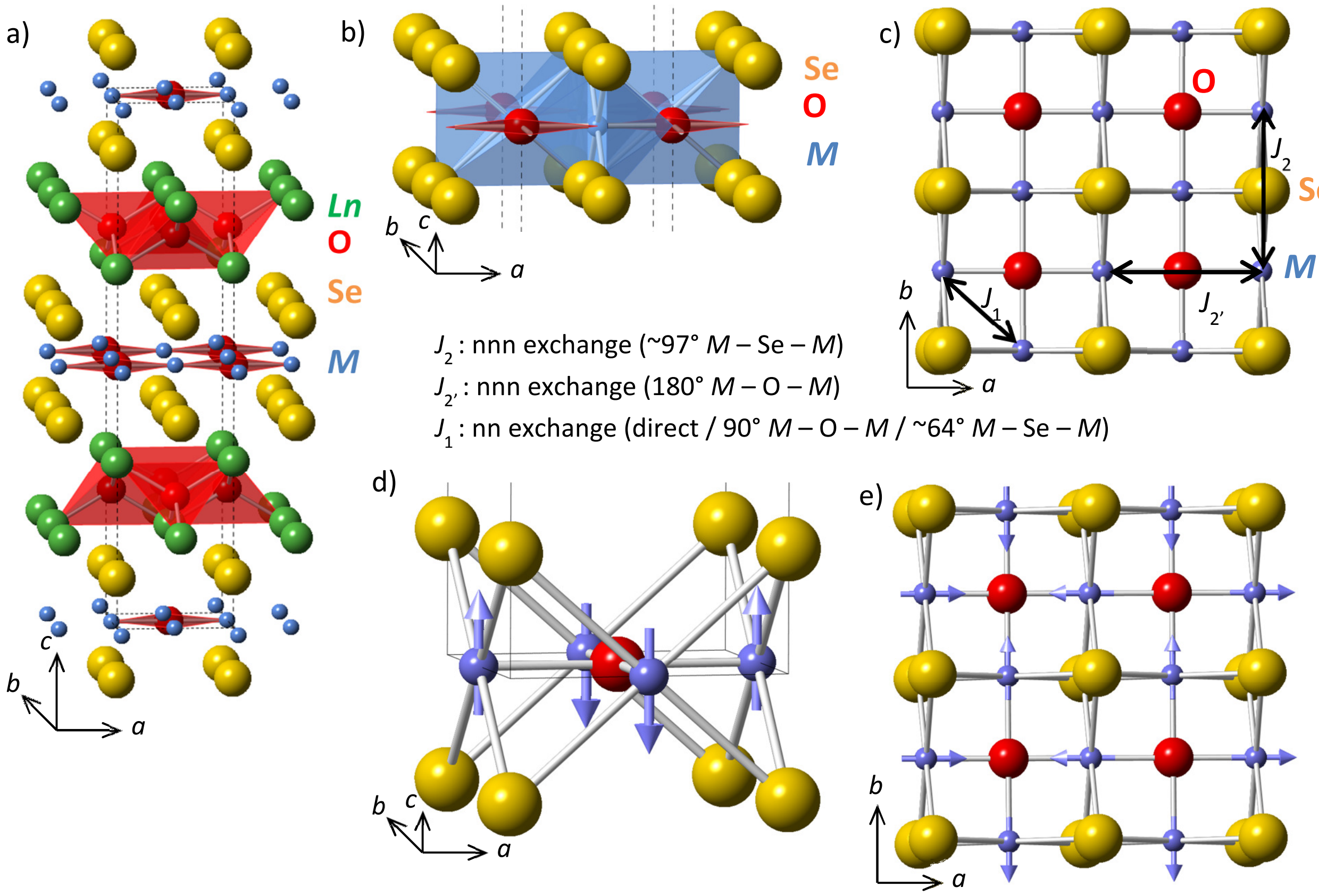

Figure 1 Structure of $Ln_2O_2M_2OSe_2$ showing (a) tetragonal unit cell, (b) pseudo-octahedral $MSe_4O_2$ coordination polyhedra and (c) *M* – Se – O layers with exchange interactions labelled; (d) shows in-plane magnetic order in $Pr_2O_2Mn_2OSe_2$[14] and (e) shows 2*k* in-plane magnetic order in $Pr_2O_2Fe_2OSe_2$, with $M^{2+}$ moments shown by blue arrows. *Ln*, *M* (*M* = Mn, Fe) O and Se ions are shown in green, blue, red and yellow, respectively.

**Experimental details**

5 g of $Pr_2O_2Fe_2OSe_2$ were prepared by the solid state reaction of $Pr_2O_3$ (prepared by heating $Pr_6O_{11}$ powder (Alfa-Aesar, 99.99%) in a flow of 5% $H_2$(g)/ 95% $N_2$(g) to 1000°C and holding at this temperature for 10 hours, before furnace cooling in the same gas flow to room temperature), Fe (Aldrich, 99.9%) and Se (Alfa-Aesar, 99.999%). Stoichiometric quantities of these reagents were intimately ground together by hand using an agate pestle and mortar. The resulting grey powder was pressed into several 5 mm diameter pellets using a uniaxial press. These pellets were slowly heated in an evacuated, sealed quartz tube to 600°C and held at this temperature for 12 hours, and then heated to 1000°C and held for 12 hours. The sample was then cooled to room temperature in the furnace. $Pr_2O_2Mn_2OSe_2$ was prepared as described elsewhere [14]. Preliminary structural characterisation was carried out using powder X-ray diffraction data collected on a Bruker D8 Advance diffractometer from 5° - 100° 2θ. The diffractometer was fitted with a LynxEye silicon strip detector (step size 0.021°) and an Oxford Cryosystems PheniX CCR cryostat to access temperatures between 12 K and 300 K. NPD data were collected for $Pr_2O_2Fe_2OSe_2$ on the TOF (time-of-flight) diffractometer WISH on target station 2 at the ISIS spallation neutron source. For scans collected in zero applied field, the powder was placed in a 6 mm diameter cylindrical vanadium can (to a height of ~6 cm). A 25 minute (15 μA h) scan was carried out at 1.5 K, then 15 minute (10 μA h) scans were collected on warming at ~1.5 K intervals up to 100 K, followed by 8 minute (5 μA h) scans at 10 K intervals up to 150 K. Data were also collected in applied magnetic fields up to 5 T by placing the pelletized sample in a 6 mm vanadium can and securing it. 25 minute (16 μA h) scans were collected every 4 K on warming from 2 K to 110 K in a 5 T applied magnetic field, and then 19 minute (12.5 μA h) scans were collected at 2 K in increments of 0.5 T up to 4.5 T. Rietveld refinements[47] were performed using TopasAcademic software.[48] Sequential refinements were carried out using data collected between 1.5 K and 150 K using TopasAcademic and local subroutines. The high resolution bank of data (bank 5, 10.5 – 85.0 ms, 1 – 4 Å) was used to investigate structural changes as a function of temperature, whilst the higher d-spacing bank 2 data (10.5 – 85.0 ms, 1 – 8 Å) were used to study the magnetic behaviour of $Pr_2O_2Fe_2OSe_2$. For sequential refinements, a background (shifted-Chebyshev polynomial, 12 terms), unit cell parameters for the nuclear phase as well as atomic coordinates, thermal displacement parameters, a peak shape for each phase and the $Fe^{2+}$ magnetic moment were refined. For data collected in applied magnetic field, three relatively strong reflections (022 at 29600 μs (d = 1.42 Å), 311 at 25200 μs (d = 1.21 Å) and 042 at 188700 μs (d = 0.90 Å)) due to aluminium (from the sample environment) were observed in the high resolution data bank and these were Pawley-fitted with a face-centred cubic model.

For INS measurements, the samples were packed into Al foil envelopes and placed in Al cans. Magnetic excitations were measured using the MARI direct geometry chopper spectrometer at the ISIS source. Incident beam energies $E_i$ = 10, 40 or 85 meV were selected using a Gd Fermi chopper set at frequencies of 250, 150 or 250 Hz, respectively. A $t_0$ chopper (spinning at 50 Hz) was used to block fast neutrons and a thick disk chopper (also spinning at 50 Hz) was used to improve the background from neutrons above the Gd absorption edge. The simulated spectra shown were obtained by convoluting the experimental resolution (measured with a vanadium standard) with a Lorentzian lineshape with terms on the energy gain and loss sides to ensure detailed balance.[49] SpinW, which uses linear spin wave theory to simulate inelastic neutron scattering on magnetic materials[50] by solving the Heisenberg Hamiltonian, was used here to similar INS spectra for various magnetic models (further details are given below).

**Results and discussion**

**1. Low temperature structural behaviour for $Pr_2O_2Fe_2OSe_2$**

NPD data collected for $Pr_2O_2Fe_2OSe_2$ could be fitted by the tetragonal structure reported by Ni et al.[36] To check our sample composition, data collected at 110 K (above $T_N$) were fitted with this model with a global temperature factor and allowing site occupancies to refine (the $Pr^{3+}$ site occupancy was fixed at unity). This gave occupancies within an esd of unity for Se, O(1) and O(2) sites, and an occupancy of 0.965(7) for the Fe site. This is very close to the ideal composition and implies formal oxidation states very close to $Pr^{3+}$, $Fe^{2+}$, $O^{2-}$ and $Se^{2-}$. The slight Fe-deficiency may imply some oxidation of $Fe^{2+}$ and $Pr^{3+}$ ions, although any oxidation must be very small given the good agreement in room temperature lattice parameters ($a$ = 4.04470(5) Å, $c$ = 18.4475(3) Å) for that expected for this series of $Ln_2O_2Fe_2OSe_2$ materials.[37]

Sequential Rietveld refinements in zero-field showed that the unit cell parameters decreased on cooling, similar to behaviour reported for other $Ln_2O_2Fe_2OSe_2$ analogues with a more rapid decrease in $c$ parameter below $T_N$, presumably due to magnetostrictive effects[17, 37]. However, a slight increase in the $c$ parameter was observed below ~22.5 K (see Figure 2b and SM1). 2 K data could be fitted by a nuclear phase of *I*4/*mmm* symmetry, and a magnetic phase (see SM2) analogous to that described for $La_2O_2Fe_2OSe_2$.[17]

Although no additional peaks and no clear splitting of diffraction peaks were observed at low temperature, we note that there is a marked increase in the d-spacing dependent peak shape term (which describes strain broadening) below ~22.5 K. Sequential refinements (using only the high resolution 153° bank data) with a Pawley phase of *I*4/*mmm* symmetry indicated that this broadening is anisotropic, affecting *h*00/0*k*0 reflections much more than *hhl* reflections (SM1). Allowing the nuclear phase to undergo an orthorhombic distortion gives stable refinements consistent with a subtle distortion to *Immm* symmetry ($a$ = 4.0390(2) Å, $b$ = 4.0349(1) Å, $c$ = 18.3820(8) Å, volume = 299.56(2) $Å^3$ at 1.5 K) involving displacement of O(2) along [001] away from the ideal 4*d* site in *I*4/*mmm* ($z$ = 0.75) (Figure 2 and SM1). NPD data collected in a 5 T applied magnetic field show that $Pr_2O_2Fe_2OSe_2$ has similar structural behaviour with and without an applied magnetic field (SM3, SM9).

With no clear peak splitting it is hard to be definitive but our analysis suggests that below ~23 K, $Pr_2O_2Fe_2OSe_2$ undergoes a distortion similar to that observed for $Pr_2O_2Mn_2OSe_2$[14] but which is more subtle for the Fe analogue (orthorhombicity, defined as $^{2(a-b)}\!/_{(a+b)}$, is 2.0 × $10^{-3}$ for $Pr_2O_2Mn_2OSe_2$ at 12 K[14], compared with 1.0 × $10^{-3}$ for $Pr_2O_2Fe_2OSe_2$ at 1.5 K (0 T)). This structural distortion may be the origin of the peak in heat capacity reported by Ni et al for $Pr_2O_2Fe_2OSe_2$ at about 23 K.[36] Analogous distortions are observed in other materials containing fluorite-like $[Pr_2O_2]^{2+}$ layers including PrMnSbO[32] (orthorhombicity ~4.3 × $10^{-3}$ at 4 K) and $PrMnAsO_{0.95}F_{0.05}$[33] (orthorhombicity 0.8 × $10^{-3}$ at 10 K). These distortions, ascribed to the $Pr^{3+}$ $4f^2$ orbital degrees of freedom[32], lower the $Pr^{3+}$ site symmetry (4*mm* to 2*mm*) as oxide ions are displaced along [001], giving two long and two short Pr–O bonds, accompanied by loss of the $C_4$ rotational symmetry and the slight increase in $c$ lattice parameter. In both PrMnSbO[32] and $PrMnAsO_{0.95}F_{0.05}$,[33] this distortion is accompanied by long range ordering of $Pr^{3+}$ magnetic moments ($T_{N,Pr}$ ≈ 80 K in PrMnSbO; $T_{N,Pr}$ ≈ 180 K in $PrMnAsO_{0.95}F_{0.05}$) within the *ab* plane as $Mn^{2+}$ moments reorientate to within this plane[32-33].

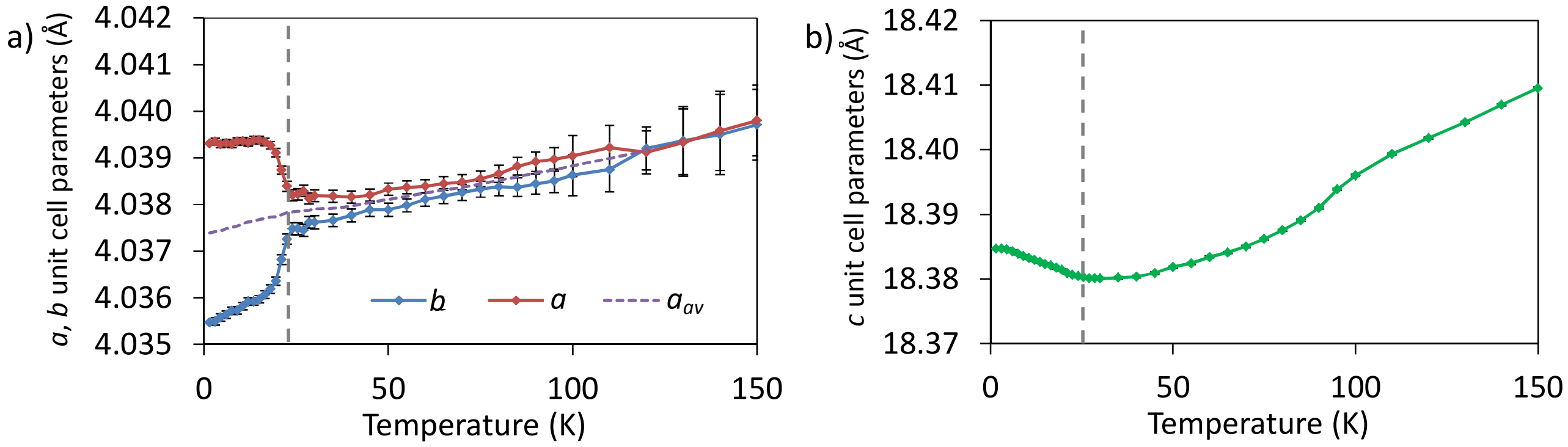

Figure 2 (Colour online) unit cell parameters from sequential NPD Rietveld refinements using *Immm* model. Dashed grey line is at 23 K and lattice parameters $a$ and $b$ are shown in red and blue, respectively, with the average in-plane lattice parameter, $a_{av}$, shown by dashed purple line.

## 2. Magnetic structure of $Pr_2O_2Fe_2OSe_2$

Ni et al showed that $Pr_2O_2Fe_2OSe_2$ orders AFM on cooling ($T_N$ = 88.6 K),[36] consistent with magnetic susceptibility measurements (SM5). In zero-field NPD data, we observed a broad asymmetric peak immediately above $T_N$ around d = 3.6 Å, from which the most intense magnetic Bragg reflection (211) grew (SM4a). The form of this peak suggests that short-range two dimensional magnetic order, that may be characterised by a Warren function,[51] occurs immediately above the transition to three-dimensional magnetic order. No Warren peak is observed above $T_N$ in 5 T applied field (SM4b).

Magnetic Bragg reflections are observed on cooling below ~90 K in zero-field. These reflections, consistent with the 2*k* magnetic structure reported for $Ln_2O_2Fe_2OSe_2$ (*Ln* = La, Nd, Ce)[17, 37] and $Sr_2F_2Fe_2OS_2$[11], increase smoothly on cooling (SM6). As for other materials, magnetic reflections are anisotropically broadened suggesting stacking faults in the magnetic structure perpendicular to the $Fe_2O$ layers. This broadening was described by an expression for antiphase boundaries perpendicular to the *c* axis,[37, 52] (SM7) with a magnetic correlation length along *c*, $\xi_c$, of 138(2) Å at 1.5 K. A good Rietveld fit was obtained with an *Immm* nuclear phase and 2*k* magnetic ordering on the Fe sites (Figure 3 and Tables 1 and 2.)

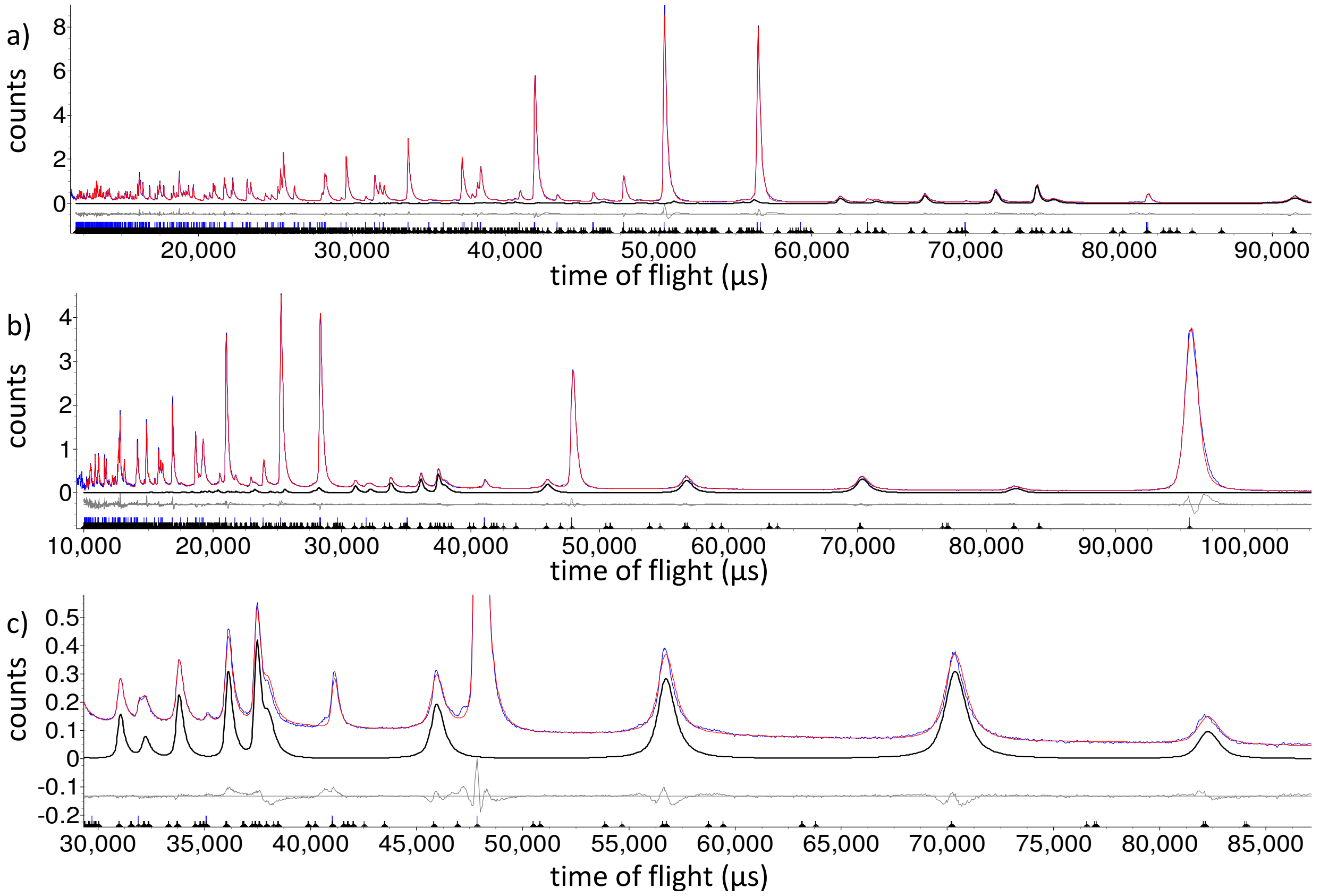

Figure 3 (Color online) Rietveld refinement profiles from combined refinement using (a) 153° bank data (~0.6 – 4.5 Å d spacing range) and (b) 59° bank data (~1 – 10 Å d spacing range) collected for $Pr_2O_2Fe_2OSe_2$ at 1.5 K; (c) highlights the higher *d*-spacing region of 59° bank data emphasising the magnetic reflections. Observed and calculated (upper) and difference profiles are shown by blue points, and red and gray lines, respectively. Magnetic intensity is highlighted by solid black line.

Table 1 Details from Rietveld refinement using NPD data collected at 1.5 K for $Pr_2O_2Fe_2OSe_2$ (zero field). The refinement was carried out with the nuclear structure described by space group *Immm* with *a* = 4.0390(2) Å, *b* = 4.0349(1) Å and *c* = 18.3820(8) Å. The magnetic scattering was fitted by a second magnetic-only phase with *a*, *b*, and *c* unit cell parameters twice those of the nuclear phase and moment orientations following Figure 1e; $R_{wp}$ = 5.07%, $R_p$ = 4.83%, and $\chi^2$ = 10.79.

| Atom | Site | *x* | *y* | *z* | $U_{iso}$ × 100 (Å$^2$) | Moment ($\mu_B$) |
|---|---|---|---|---|---|---|
| Pr | 4*i* | 0 | 0 | 0.68612(6) | 0.35(4) | |
| Fe(1) | 2*d* | 0.5 | 0 | 0.5 | 0.18(1)* | 3.36(1) |
| Fe(2) | 2*b* | 0 | 0.5 | 0.5 | 0.18(1)* | 3.36(1) |
| Se | 4*i* | 0 | 0 | 0.09806(4) | 0.03(2) | |
| O(1) | 4*j* | 0.5 | 0 | 0.7523(3) | 0.23(3) | |
| O(2) | 2*c* | 0.5 | 0.5 | 0 | 0.52(4) | |

* $U_{iso}$ for Fe(1) and Fe(2) constrained to be the same.

Table 2 Selected bond distances from Rietveld refinement using NPD data collected at 1.5 K for $Pr_2O_2Fe_2OSe_2$ (zero field).

| **Bond** | **Length (Å)** |
|---|---|
| Pr – O(1) | 2 × 2.313(2) |
| Pr – O(1) | 2 × 2.358(3) |
| Fe(1) – O(2) | 2 × 2.0195(1) |
| Fe(1) – Se | 4 × 2.7054(4) |
| Fe(2) – O(2) | 2 × 2.01743(8) |
| Fe(2) - Se | 2 × 2.7069(4) |

Attempts to include an ordered moment on the Pr site (in zero field) in $Pr_2O_2Fe_2OSe_2$ gave no improvement in fit ($R_{wp}$ decreased by 0.0003% for this additional parameter) and a $Pr^{3+}$ moment of zero within two standard uncertainties (0.11(8) $\mu_B$); we conclude that there is no long-range order of $Pr^{3+}$ moments at 1.5 K in zero-field. We note that the $Fe^{2+}$ moment determined here is larger than that reported by Ni et al (2.23(3) $\mu_B$ at 5 K)[36] (presumably due to improved fitting of broadened magnetic Bragg reflections), but is similar to that reported for other $Fe_2O$ phases.[11, 17, 37, 52] Sequential Rietveld refinements carried out using the 59° data bank indicate that the onset of the 2*k* magnetic order on the Fe sublattice (Figure 4) is 2D-Ising-like, similar to other $Fe_2O$ phases.[17, 25, 37] The higher order terms needed to stabilise the 2*k* structure couple the perpendicular *k*-vectors and introduce the $C_4$ rotational symmetry to the $Ln_2O_2Fe_2OSe_2$ magnetic symmetry.[17] These terms may compete with the $Pr^{3+}$-driven orthorhombic distortion, but are likely to be much smaller in energy than the Pr orbital degrees of freedom that drive the structural distortion.

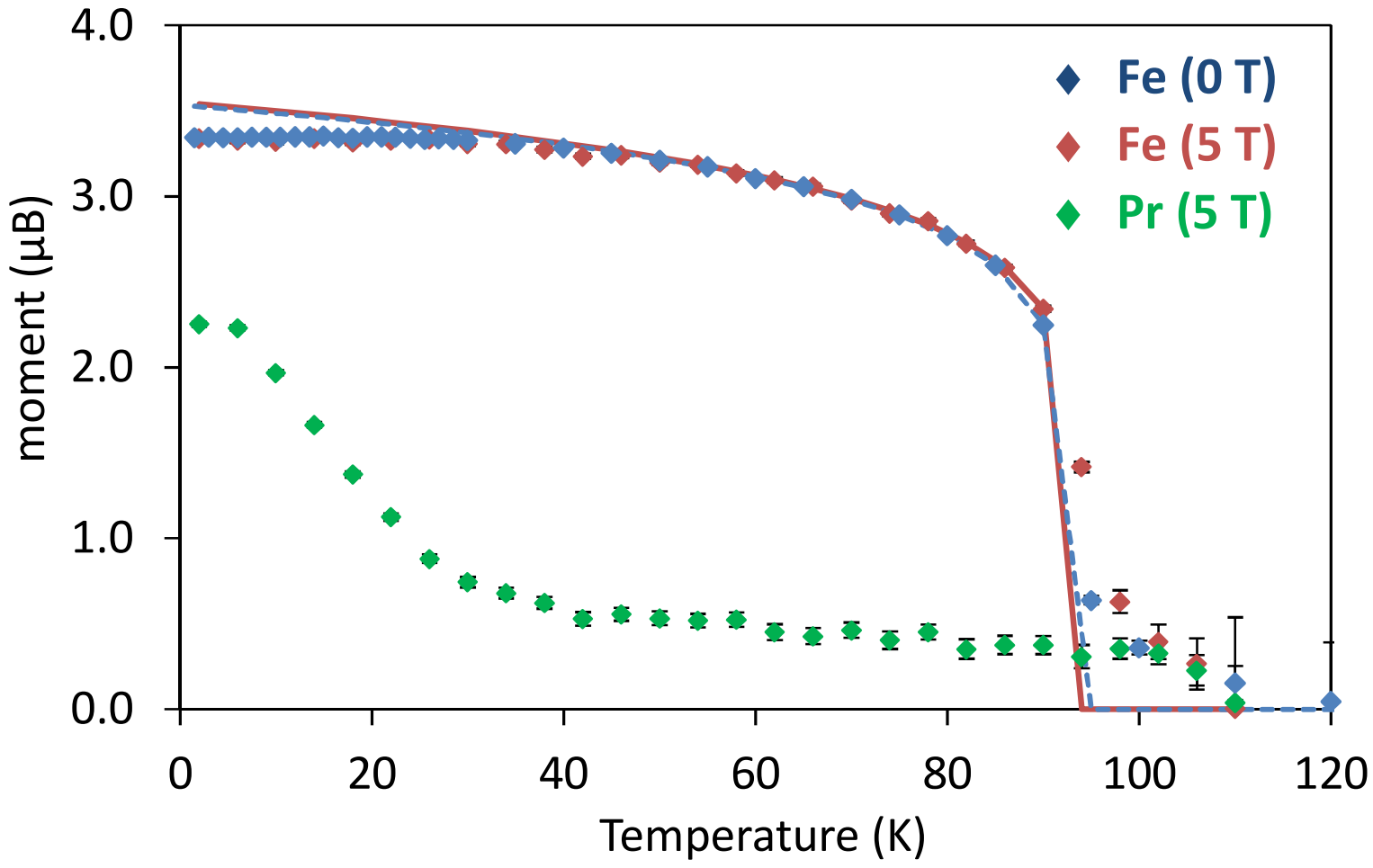

Figure 4 (Color online) Evolution of magnetic moments on Fe sites in zero applied magnetic field (blue diamonds), and in 5 T (red diamonds) and on Pr sites in 5 T (green diamonds) from sequential Rietveld refinements; dashed blue (zero field) and solid red (5 T) lines are guides to the eye showing critical behaviour for a 2D Ising-like system (with critical exponent $\beta \approx 0.1(1)$).

Ni et al suggested that $Pr_2O_2Fe_2OSe_2$ might undergo a metamagnetic phase transition at low temperatures,[36] consistent with the field-dependence observed in our susceptibility measurements (SM5). This prompted us to use NPD to study the effect of an applied magnetic field. In 5 T applied field, magnetic Bragg reflections (consistent with the 2*k* magnetic structure) are observed at 94 K and increase in intensity on cooling (SM8). Below ~26 K in 5 T field, additional magnetic scattering is observed at *k* = (0 0 0) positions (relative to the nuclear cell). These reflections are sharper than those due to the 2*k* ordering (Figure 5). Attempts to fit the low temperature 5 T NPD data with various ordering patterns on the Fe sublattice were unsuccessful but including a FM component on $Pr^{3+}$ sites did improve the fit. 1.5 K NPD data collected in 5 T field were fitted with 2*k* magnetic ordering on the Fe sublattice (with anisotropic peak broadening of these reflections as described above), and with $Pr^{3+}$ moments along the shorter *b* axis (SM9). The same peakshape was used to fit nuclear reflections and the additional reflections arising from the FM component, suggesting no stacking faults for this field-induced FM component. The $Pr^{3+}$ moment at 5 T (2.21(1) $\mu_B$) is smaller than that found in PrMnSbO (2.96(3) $\mu_B$) and $PrMnAsO_{0.95}F_{0.05}$ (~3 $\mu_B$)[33] but larger than that found in the poorly-metallic PrFeAsO (0.83(9) $\mu_B$)[53].

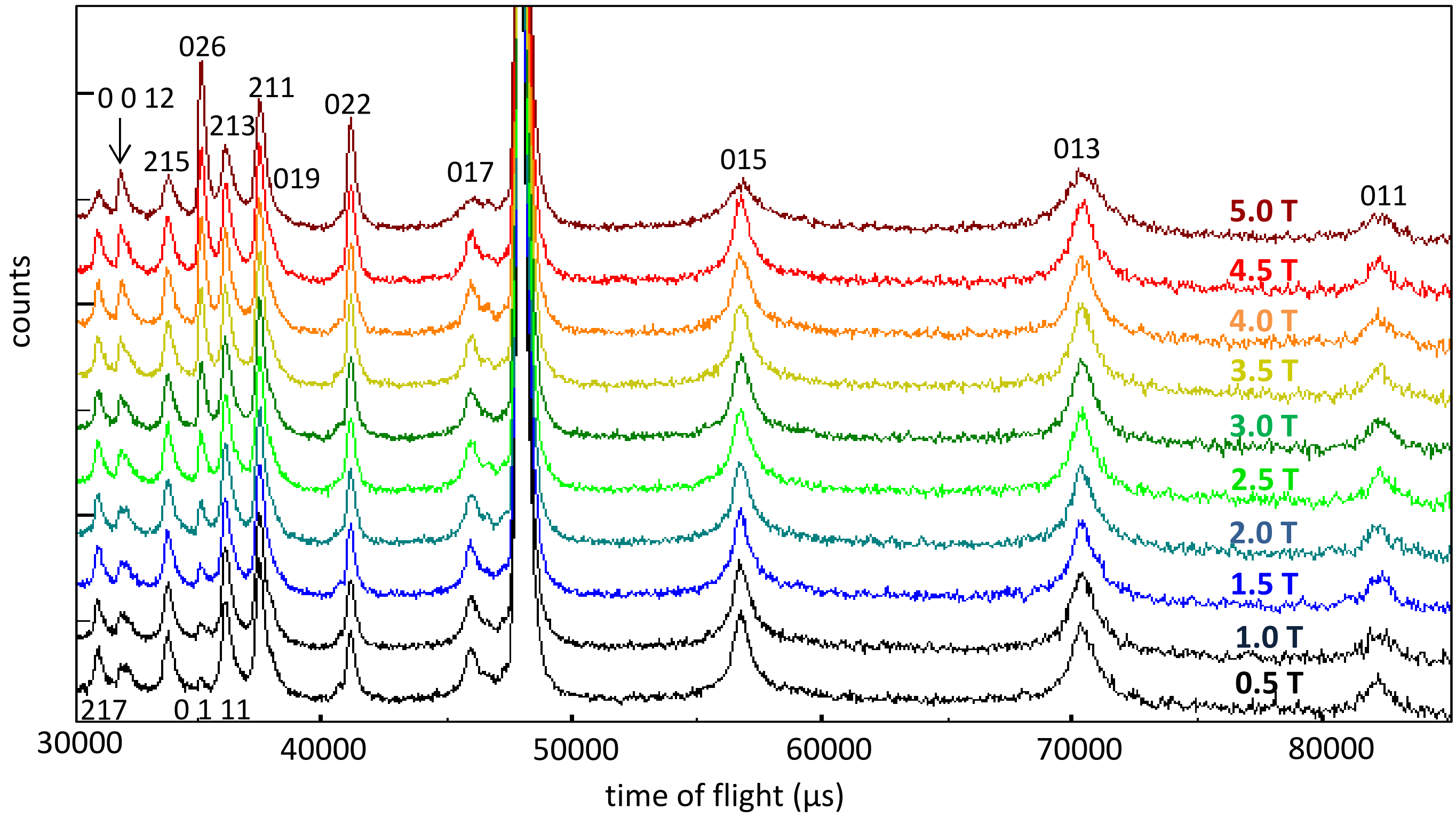

Figure 5 (Color online) 59° bank NPD data collected for $Pr_2O_2Fe_2OSe_2$ at 2 K in applied magnetic field showing evolution of new magnetic Bragg reflections (022, 026) as a function of applied field, as well as broadening of existing 2-*k* magnetic reflections (e.g. 015, 017).

Whilst this model gives a good fit to our NPD data, it may not correspond to the ground state magnetic structure in an applied magnetic field due to the difficulties associated with working with powder samples and significant spin anisotropy. Although crystallites in our pelletized sample are unlikely to reorient in the applied (uniaxial) field, the powder averaging inherent in a "powder diffraction" experiment is likely to break down: the applied magnetic field will cause moments to reorient depending on the relative direction of the applied field, orientation of the crystallite and direction of any spin anisotropy (particularly relevant to the $Pr^{3+}$ sites). A single crystal sample would be needed to overcome these effects.

### 3. Transition metal magnetic excitations in $Pr_2O_2Fe_2OSe_2$ and $Pr_2O_2Mn_2OSe_2$

In this section we discuss the excitations on the transition metal site in $Pr_2O_2Fe_2OSe_2$ and the manganese analogue $Pr_2O_2Mn_2OSe_2$ (which shows similar but more marked orthorhombicity at low temperatures). In the following section we will present the localized excitations from the crystalline electric field on the $Pr^{3+}$ site. The magnetic excitations for $Pr_2O_2Fe_2OSe_2$ and $Pr_2O_2Mn_2OSe_2$ are shown in Figure 6 measured on the MARI spectrometer at 5 K with $E_i$ = 40 meV (the same configuration used to study $La_2O_2Fe_2OSe_2$[17]). The large momentum-dependence of the intensity (Figures 6a, c) indicates that it is due to magnetic excitations associated with the $Fe^{2+}$ or $Mn^{2+}$ sites rather than the $Pr^{3+}$ site. As discussed for $Ce_2O_2FeSe_2$,[24] low energy transition metal and crystal field excitations can be distinguished by their momentum dependence, with excitations associated with the lanthanide site ($Pr^{3+}$ in this case), local crystal field excitations, showing no strong momentum dependence.

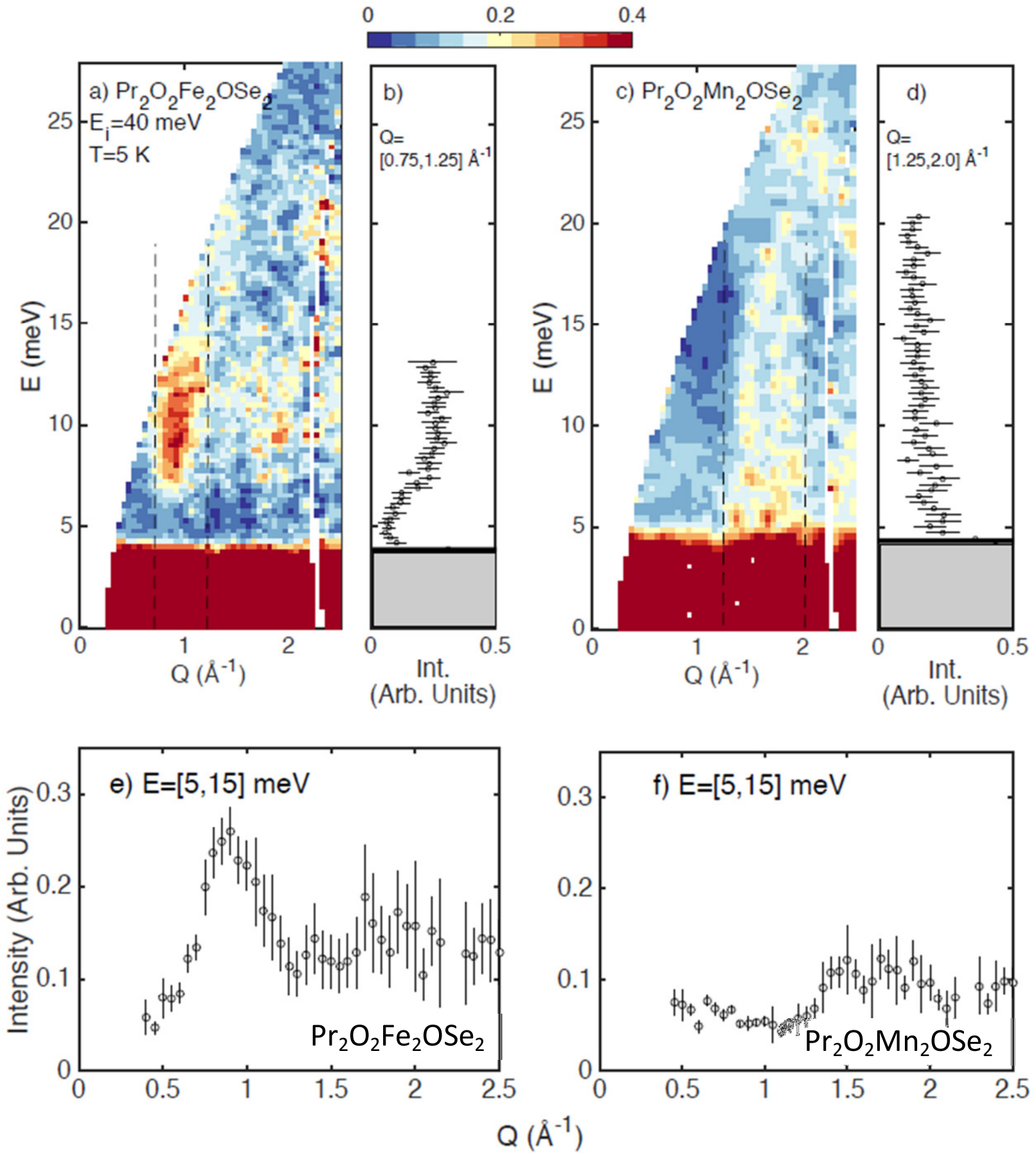

Figure 6 (a) Energy momentum slice for $Pr_2O_2Fe_2OSe_2$ at 5 K, (b) a constant momentum cut integrating over $Q$ = [0.75, 1.25] Å$^{-1}$, (c) energy momentum slice for $Pr_2O_2Mn_2OSe_2$ at 5 K, (d) a constant momentum cut integrating over $Q$ = [1.25, 2.00] Å$^{-1}$, constant energy slices are shown for $Pr_2O_2Fe_2OSe_2$ (e) and for $Pr_2O_2Mn_2OSe_2$ (f) integrating over [5,15] meV. The area contaminated by incoherent scattering from the $E$ = 0 elastic line is shaded in grey in panels (b) and (d).

We consider first $Pr_2O_2Fe_2OSe_2$ (Figure 6a, b and e). The magnetic cross section (Figure 6a) is almost identical to that observed for $La_2O_2Fe_2OSe_2$[17] (further supporting our assignment of the fluctuations to the $Fe^{2+}$ site rather than the $Pr^{3+}$ site). The low-energy excitation spectrum is dominated by an anisotropy gap. Due to powder averaging and the two-dimensional nature of the magnetic fluctuations, the gapped excitations manifest as a broadened "step" in the excitation spectrum rather than a sharp peak. This has been observed for other systems including the sum rule analysis of neutron scattering data for $La_2O_2Fe_2OSe_2$[17] (see Figure 3d of reference 17) as well as for low dimensional magnets α-$NaMnO_2$ (see Figure 2 of reference 54) and Cu(quinoxaline)$Br_2$ (see Figure 7 of reference 55). The similar magnetic cross sections for $Pr_2O_2Fe_2OSe_2$ and $La_2O_2Fe_2OSe_2$ also suggests that $Pr_2O_2Fe_2OSe_2$ has similar exchange interactions to $La_2O_2Fe_2OSe_2$, which were modelled in terms of dominant AFM Fe – O – Fe $J_{2'}$ interactions and weaker FM Fe – Se – Fe $J_2$ interactions (spectra compared against calculations with $J_1$ = 0.75 meV, $J_2$ = -0.10 meV and $J_{2'}$ = 1.00 meV). A large anisotropy results in a gap in the momentum slice (Figure 6a) and this is confirmed in the momentum integrating cut (Figure 6b) which shows an onset of magnetic intensity at ~5 – 6 meV. The similar magnetic behaviour of the iron sublattice in both La and Pr analogues highlights the limited influence of the lanthanide ion (size, and magnetic properties) on the robust 2*k* magnetic order on the iron sublattice.[37]

The magnetic response in $Pr_2O_2Mn_2OSe_2$ (Figure 6c, d and f) differs from the $Fe^{2+}$ analogue, displaying a more complex momentum dependence consistent with several strong and competing exchange interactions within the Mn sublattice.[14-15] This is consistent with the frustrated nature of $Mn_2O$ materials suggested by NPD analysis and property measurements.[13, 15] The bulk of the spectral weight in the $Mn^{2+}$ analogue is shifted to larger momentum transfers (Figure 6f) indicating an exchange interaction over a smaller length scale becoming dominant.[56] There is also significant spectral weight in the limit Q → 0 which may indicate multiple competing exchange interactions corresponding to different lengthscales. On the scale of the resolution (FWHM ≈ 2.2 meV), no anisotropy gap is observed for $Pr_2O_2Mn_2OSe_2$ and the momentum integrating cut (Figure 6d) is smooth and almost independent of energy, in contrast with that for $Pr_2O_2Fe_2OSe_2$ (Figure 6b). This is consistent with the more three-dimensional character of the onset of $Mn^{2+}$ magnetic ordering suggested by NPD analysis (critical exponent β = 0.24(3) for $La_2O_2Mn_2OSe_2$, compared with 0.122(1) for $La_2O_2Fe_2OSe_2$).[14, 17]

### 3.1 Linear spin wave analysis to determine exchange interactions in $Pr_2O_2Mn_2OSe_2$

To compare these $Fe^{2+}$ and $Mn^{2+}$ oxyselenides, linear spin wave calculations were carried out with the aim of reproducing the features which distinguish the $Mn^{2+}$ phase from the $Fe^{2+}$ analogue. The calculations were done using the SpinW package, based on the Heisenberg Hamiltonian

$$\mathcal{H} = \sum_{ij} J_{ij} \vec{S_i} \cdot \vec{S_j} \quad (1)$$

where *i*, *j* are magnetic sites and $J_{ij}$ is the coupling between them ($J < 0$ indicates FM exchange whilst $J > 0$ indicates AFM exchange). These calculations, to investigate the magnetic excitations from the $Mn^{2+}$ sublattice, were performed on $La_2O_2Mn_2OSe_2$ (excitations from the $Pr^{3+}$ sublattice are not relevant for this analysis, as explained above).

Given the complex neutron response and the number of exchange interactions, general fits to the inelastic spectra were not possible; we therefore investigate the effect of each exchange parameter on the neutron response (Figure 7). A dominant $J_1$ interaction reproduces the "wall" of scattering at larger momentum transfers (Figure 7a), consistent with the experimentally observed magnetic structure.[14-15] The tuning of the nnn exchanges $J_2$ and $J_{2'}$ gives a shift of the magnetic spectral weight which is inconsistent with the observed spectra (Figure 6c). From this, we conclude that the strong AFM nn exchange $J_1$ is dominant.

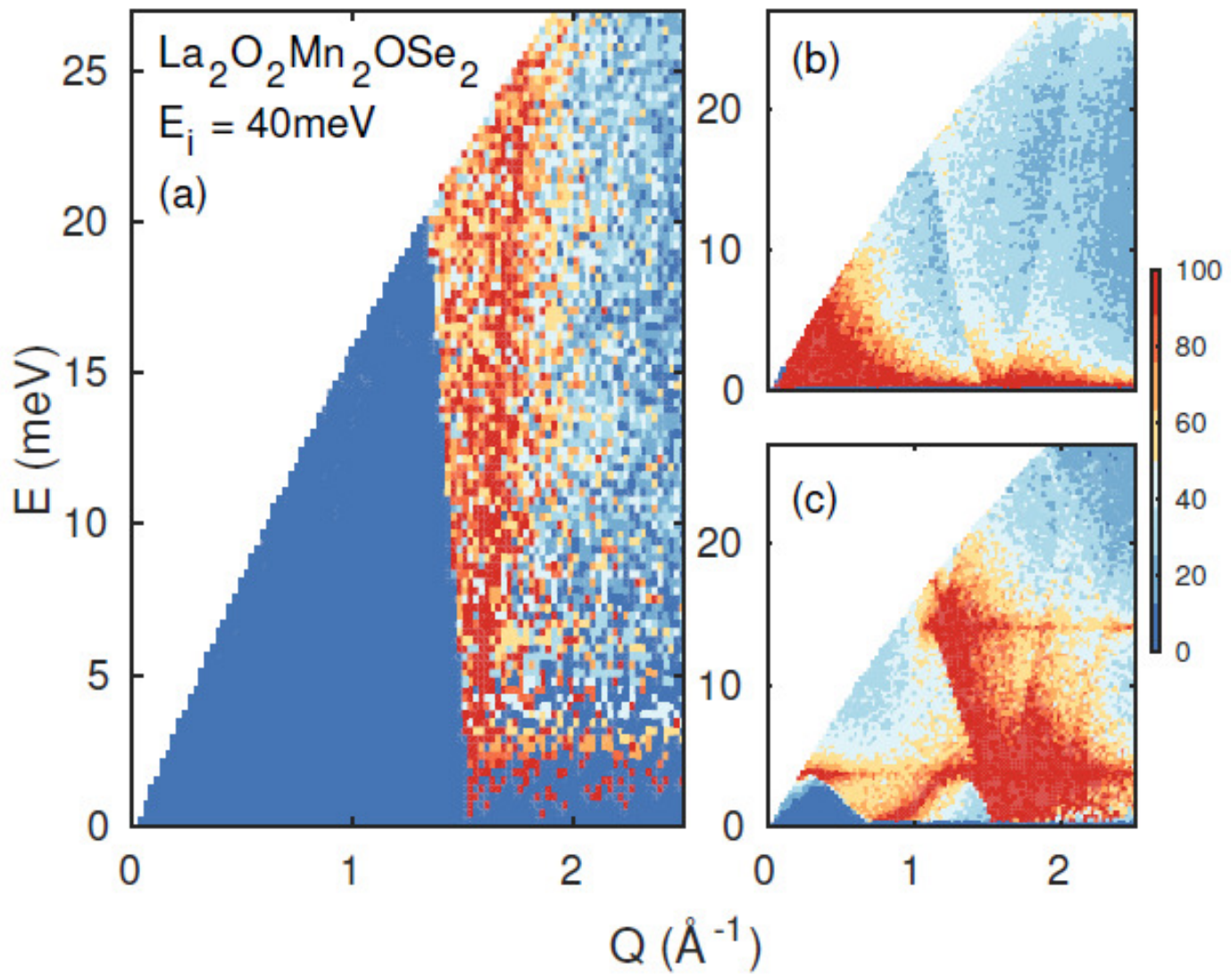

Figure 7 Simulated INS powder spectra for $La_2O_2Mn_2OSe_2$ using SpinW with exchange constants (a) $J_1$ = 5.05 meV and $J_2$ = $J_{2'}$ = 0; (b) $J_2$ = -2.8 meV and $J_1$ = $J_{2'}$ = 0 and (c) $J_1$ = $J_{2'}$ = 1.4 meV and $J_2$ = -1.4 meV.

The effect of tuning nnn interactions $J_2$ (Mn – Se – Mn) and $J_{2'}$ (Mn – O – Mn) with fixed AFM $J_1$ are investigated in Figure 8. There is some ambiguity from theoretical work as to the sign of the nnn Mn–Se–Mn $J_2$ interaction,[13, 57] but our model considers FM $J_2$ exchange, reflecting the experimental magnetic structure. Figure 8 (a – c) illustrates the effect of increasing the FM $J_2$ on the neutron response: little qualitative change is observed and we can't determine the value of FM $J_2$ from this analysis. Panels (d – f) illustrate the effect of introducing AFM $J_{2'}$ (180° Mn – O – Mn exchange) on the neutron response and we note that it is in competition with AFM nn $J_1$: both these AFM interactions cannot be satisfied simultaneously. On introducing $J_{2'}$ (Figure 8 d – f), we see a shift in spectral weight to lower energies and to lower momentum transfers. While this may be consistent with the presence of spectral weight in the limit Q →0 (Figure 6f), $J_{2'}$ does seem to reduce the "wall" of scattering at larger momentum transfers. Based on this comparison of our experimental data with linear spin wave calculations, we estimate values for the exchange interactions ($J_1$ ≈ 5 meV, $J_2$ ≈ -5 meV and $J_{2'}$ ≈ 1-5 meV ) which are consistent with the higher degree of magnetic frustration for the Mn (compared with for example, *M* = Fe systems) expected from analysis of diffraction data and heat capacity measurements.[13, 15]

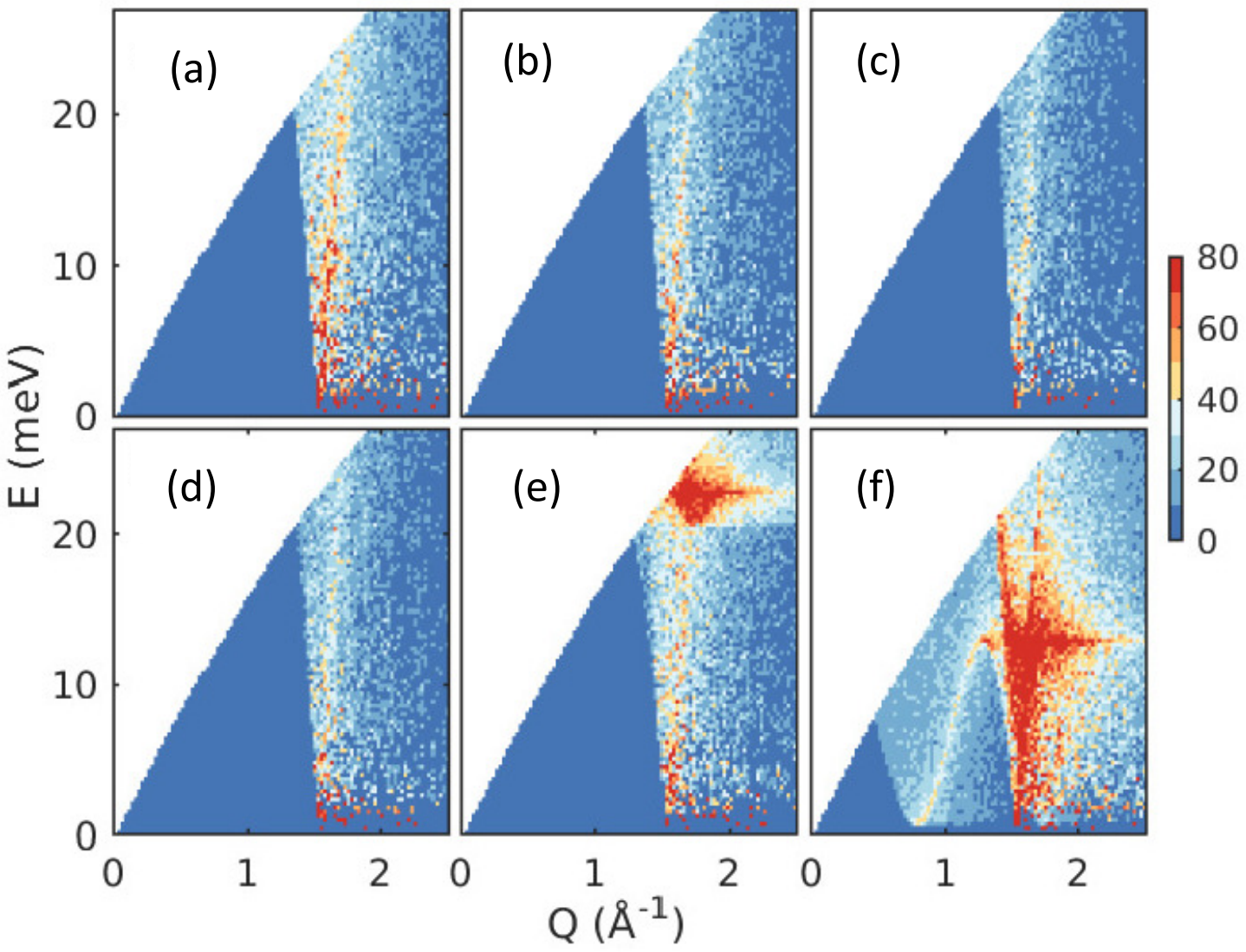

Figure 8 SpinW simulated INS spectra simulated for $La_2O_2Mn_2OSe_2$ with fixed nn $J_1$ = 5.05 meV. Top row has fixed $J_{2'}$ = 0 and $J_2$ = -1, -3 and -5 meV for (a), (b) and (c), respectively. Bottom row has fixed $J_2$ = -5 meV and $J_{2'}$ = 1, 3 and 5 meV for (d), (e) and (f), respectively.

## 4. $Pr^{3+}$ crystal field excitations

Having discussed excitations associated with the transition metal sublattice, we now discuss the local electronic crystal field probed through excitations on the $Pr^{3+}$ site. Given the local symmetry of the $Pr^{3+}$ site (the 4*e* site of $C_{4v}$ (4*mm*) symmetry) in the tetragonal *I*4/*mmm* phase, the Hamiltonian associated with the crystalline electric field takes the form described by Stevens operators[58]:

$$H_{tet} = B_2^0O_2^0 + B_2^2O_2^2 + B_4^0O_4^0 + B_4^4O_4^4 + B_6^0O_6^0 + B_6^4O_6^4 \quad (2)$$

The orthorhombic distortion lowers the symmetry of the $Pr^{3+}$ site (to $C_{2v}$ (*mm*2) symmetry at the 4*i* site of the orthorhombic *Immm* model) and additional terms are needed in the Hamiltonian to describe the crystal field:

$$H_{ortho} = B_2^0O_2^0 + B_2^2O_2^2 + B_4^0O_4^0 + B_4^2O_4^2 + B_4^4O_4^4 + B_6^0O_6^0 + B_6^2O_6^2 + B_6^4O_6^4 + B_6^6O_6^6 \quad (3)$$

We now discuss how these terms affect the neutron scattering cross section. In the dipole approximation for localized magnetic moments, the neutron scattering cross section at small momentum transfers can be written[59]

$$\frac{d^2\sigma}{d\Omega d\omega} = \frac{(\gamma r_0)^2}{4}\frac{k_f}{k_i}f^2(Q)\sum_{r,s}\rho_n|\langle r|J_\perp|s\rangle|^2\delta(E_r - E_s - \hbar\omega) \quad (4)$$

where $\frac{(\gamma r_0)^2}{4}$ is 73 mbarns sr$^{-1}$, $f^2(Q)$ is the magnetic form factor, the indices *r,s* refer to the different crystalline field wavefunctions, $\rho_n$ is a matrix element weighting factor and $J_\perp$ is the projection of *J* perpendicular to *Q*. This cross section illustrates a selection rule $\Delta m = 0, \pm1$ (where $m$ are the eigenvalues of the $J_\perp$ operator). (Magnetic crystal field excitations are distinguished from collective excitations associated with exchange interactions between magnetic $Fe^{2+}$ ions by their momentum dependence, as discussed above and for $Ce_2O_2FeSe_2$.[24])

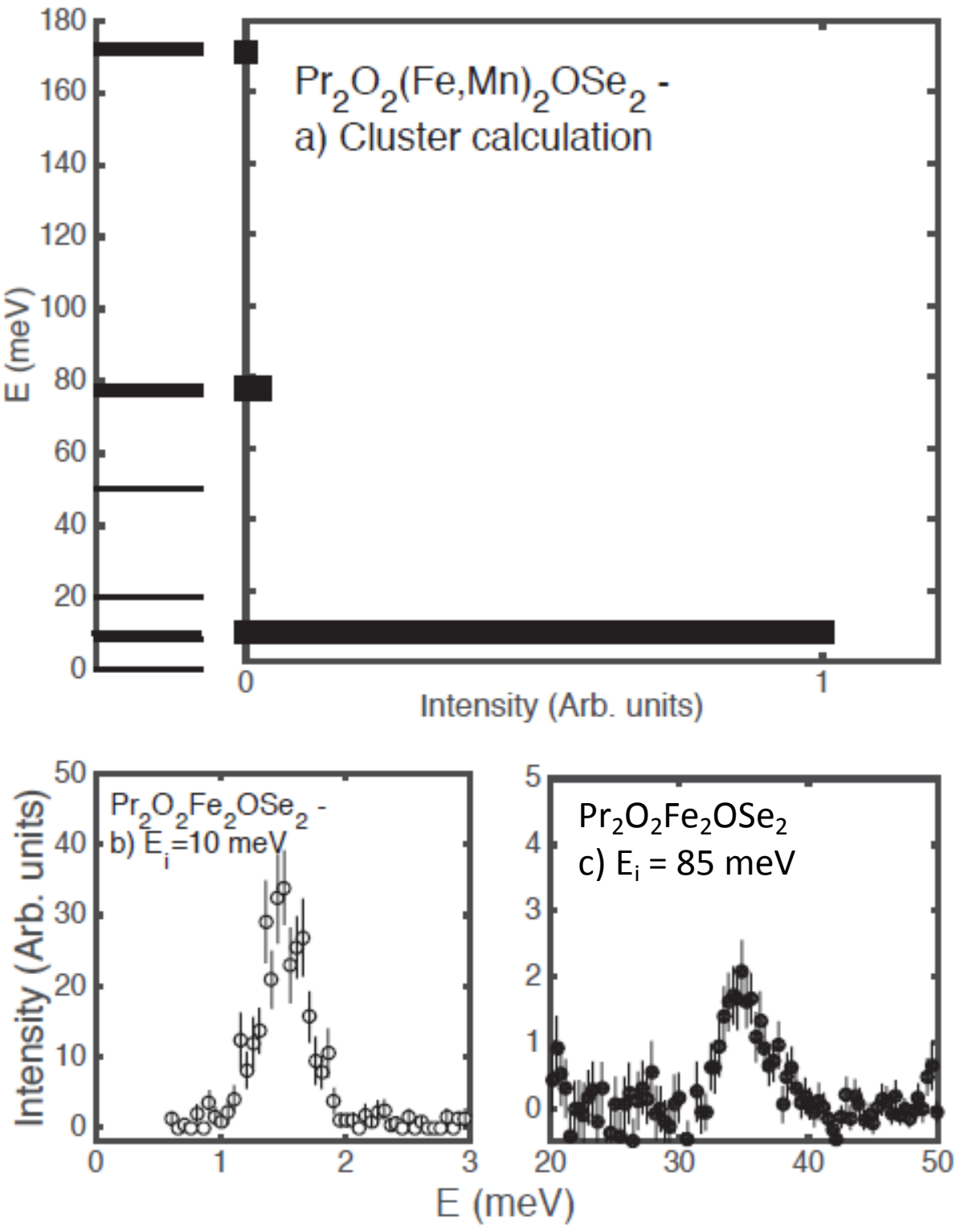

Figure 9 (a) Calculated crystal field scheme for $Pr^{3+}$ ions in $Pr_2O_2M_2OSe_2$ ($M$ = Mn, Fe) assuming a point charge model summing over 10 tetragonal unit cells. The calculation shows three allowed transitions, two of which are observed and are shown in (b) and (c) which show energy scans integrating in momentum over $Q$ = [0,5] $Å^{-1}$) for $Pr_2O_2Fe_2OSe_2$. The third, high energy, crystal field level was not observed experimentally, consistent with the weak cross section expected from the point charge calculation.

We have performed a point charge calculation for the crystal field Hamiltonian with $Pr^{3+}$ ($J$ = 4) outlined in equation 2. The results of this "cluster" point charge calculation are shown in Figure 9 with energy positions and calculated neutron scattering dipole intensities displayed. The calculation was performed considering localised point charges with Stevens parameters obtained by summing over ten unit cells to ensure convergence.[60] This approximate approach is justified by the fact that Stevens parameters depend on sums that decay rapidly with radius $R$ as $\sim {}^{1}\!/_{R^{(L+1)(2L+1)}}$ with $L$ = 2, 4, 6, facilitating a rapid convergence. These calculations give similar results to those on tetragonal $Pr_2CuO_4$ with a singlet ground state and a first excited state with a large calculated and measured neutron cross section.[61-63] We note that while the crystal field levels are either singly or doubly degenerate, the wavefunction is a linear combination of the nine eigenstates of $J_z$ associated with a total angular momentum $J$ = 4. While the point charge model is not entirely appropriate for these oxyselenides (in which covalency is likely to be important), it does provide a useful starting point for understanding the properties of the crystal field excitations and the degeneracy.

The results show that the lowest energy crystal field excitation from the singlet ground state is a doublet, and this excitation is calculated to have the largest neutron scattering intensity. Two other excitations were calculated to have finite neutron cross sections and are also illustrated in Figure 9, though with considerably weaker intensities. $Pr^{3+}$ is a non-Kramers system with $J$ = 4, therefore the splitting of the excitations is sensitive to the crystal field local symmetry. In particular, on including terms in the Hamiltonian listed in equation 3 for an orthorhombic unit cell, splitting of the doublet crystal field excitations is recovered. This is in contrast to the Kramers system involving $Ce^{3+}$ where Kramers theorem guarantees that the excitation spectrum consists of doublets that cannot be split unless a magnetic field is applied, as is the case for a magnetically ordered system, as demonstrated by neutron scattering studies for CeFeAsO.[41] On inclusion of terms allowed for orthorhombic symmetry (terms included in $H_{ortho}$ , eq. 3), this doublet splits, making this lowest energy excitation sensitive to local electrical inhomogeneity of local distortions.

The lowest two crystal field excitations for $Pr^{3+}$ in $Pr_2O_2Fe_2OSe_2$ are shown in Figure 9b, c. Only two crystal field excitations are observed in this energy range, consistent with the point charge "cluster" calculations described above. The most intense excitation occurs at low energies (~1.5 meV) and the next highest excitation occurs at ~35 meV. These two excitations have intensities similar to those predicted from the point charge cluster calculations, but appear at significantly lower energies than predicted by the calculation. This difference in observed and calculated energies is presumably due to orbital overlap and covalency which requires a more detailed electronic theory than that afforded by the point charge model. The crystal field energies, level scheme and neutron cross sections measured here for $Pr_2O_2Fe_2OSe_2$ are consistent with those for $Pr^{3+}$ in cuprate systems[28-30] including $Pr_2CuO_4$.

Given the strong neutron scattering intensity of the lowest energy $Pr^{3+}$ excited doublet, and its sensitivity to local ligand fields, we studied this excitation as a function of temperature using the MARI spectrometer with $E_i$ = 10 meV (Figure 10). Panel 10a illustrates a resolution-limited excitation at low temperatures (3.5 K) which broadens on warming (Figure 10b and 10c at 40 K and 75 K, respectively). The linewidth and energy position are shown in Figure 10d and e. Both parameters show a change at 30 – 50 K and then a saturation below ~ 20 K. At the lowest temperatures, the first crystal field excitation of $Pr_2O_2Fe_2OSe_2$ is resolution-limited and no splitting (from the subtle crystallographic orthorhombic distortion) is observed.

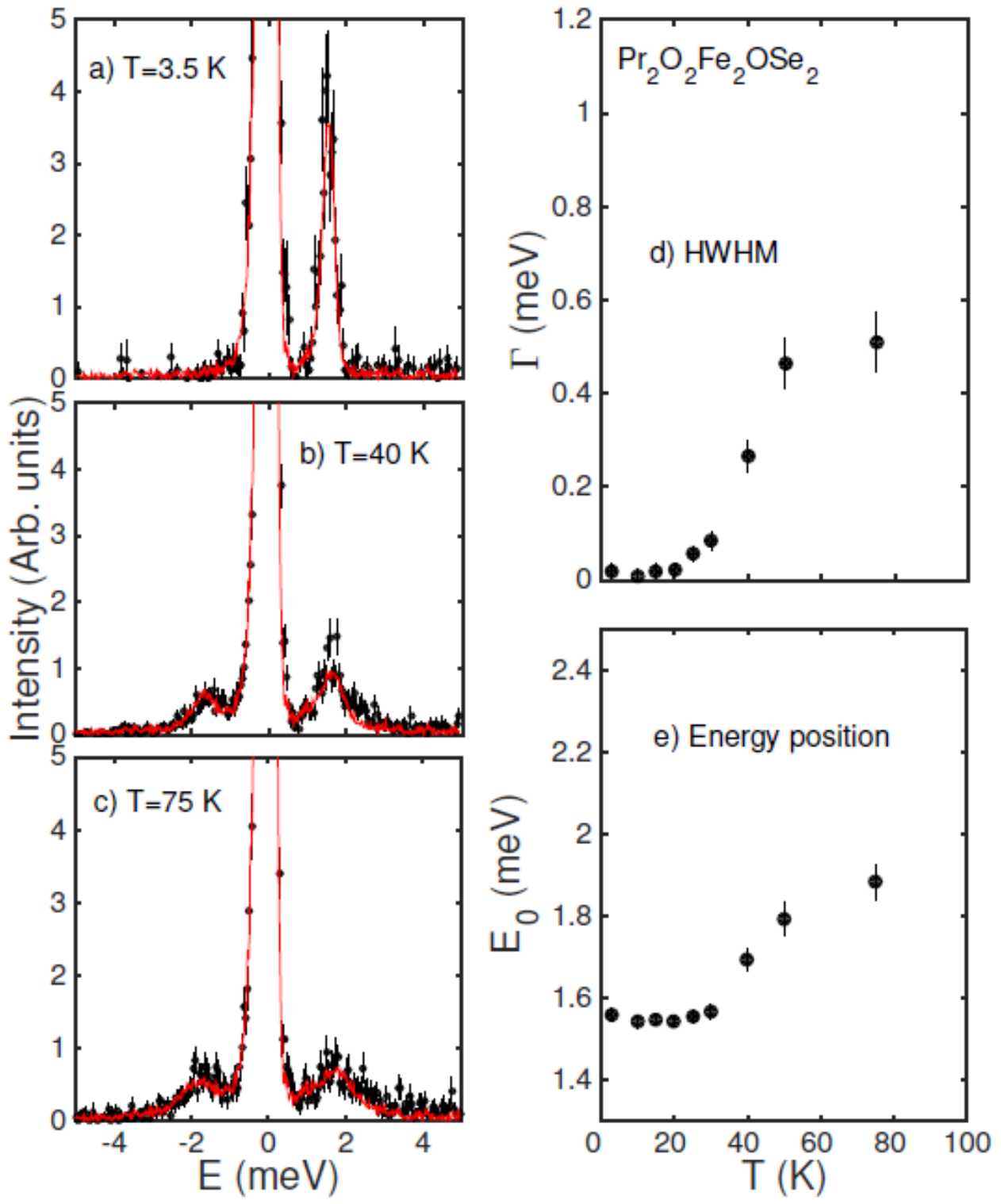

Figure 10 The lowest energy crystal field excitation in $Pr_2O_2Fe_2OSe_2$: the linewidth and energy of the excitation are shown as a function of temperature (panels (d) and (e), respectively).

The effects of an applied magnetic field on $Pr_2O_2Fe_2OSe_2$ were investigated using this point charge "cluster" model described above by adding a Zeeman term to the Hamiltonian given in equation 3. A weak transverse magnetic field splits the $Pr^{3+}$ excited doublet and results in a weak $Pr^{3+}$ transverse moment (typically ~ 1 $\mu_B$ per $Pr^{3+}$ site for a 5 T field). Calculations with the external field perpendicular to the (001) planes did not give any such mixing and therefore resulted in zero $Pr^{3+}$ moment. The experimental observation of an ordered in-plane $Pr^{3+}$ moment in an applied magnetic field for $Pr_2O_2Fe_2OSe_2$ (as described above in 5 T field, see Figures 4 – 5 and SM9) is consistent with our calculations and CEF understanding.

The temperature dependence of the low energy $Pr^{3+}$ crystal fields for $Pr_2O_2Mn_2OSe_2$ are shown in Figure 11. In contrast to the $Fe^{2+}$ analogue discussed above, at 3.5 K the low energy excitation (Figure 11a) shows two distinct components, with a sharp component at ~2.8 meV and an overdamped component at E = 0. The excitation was fitted using two Lorentzian peaks at low temperatures. At high temperatures, only one temporally overdamped mode is observed and so only one overdamped Lorentzian term was used to fit these higher temperature data. The linewidth as a function of temperature is shown in Figure 11d for the temperature range for which two Lorentzian terms were used. The overdamped component is observed even at the lowest temperatures (Figure 11d), in contrast to the $Fe^{2+}$ analogue for which the temporally sharp (resolution-limited in energy) crystal field excitations are observed at low temperatures. From our data for $Pr_2O_2Mn_2OSe_2$, it is not possible to determine whether the overdamped component is due to splitting of the first excited doublet, or due to a broadening in energy of the lowest energy ground state. The broad excitation observed for $Pr_2O_2Mn_2OSe_2$ may allow a non-zero static moment on $Pr^{3+}$ sites through mixing of the lowest energy crystal fields. It's interesting that there's no evidence for long range $Pr^{3+}$ magnetic order in $Pr_2O_2M_2OSe_2$ whilst $Pr^{3+}$ moments do order in several related 1111 materials including PrFeAsO[31], PrMnAsO[33] and PrMnSbO[32], in which the $Pr^{3+}$ order is often coupled to the transition metal magnetism. A static $Pr^{3+}$ moment (~3 $\mu_B$ at 2 K) was observed in $PrMnAsO_{0.95}F_{0.05}$ [33] which has a smaller orthorhombic distortion; we can't rule out the possibility of a small localised moment on $Pr^{3+}$ sites in $Pr_2O_2Mn_2OSe_2$ that our NPD data aren't sensitive to and further studies, for example by muon spin relaxation (μSR), would be of interest.

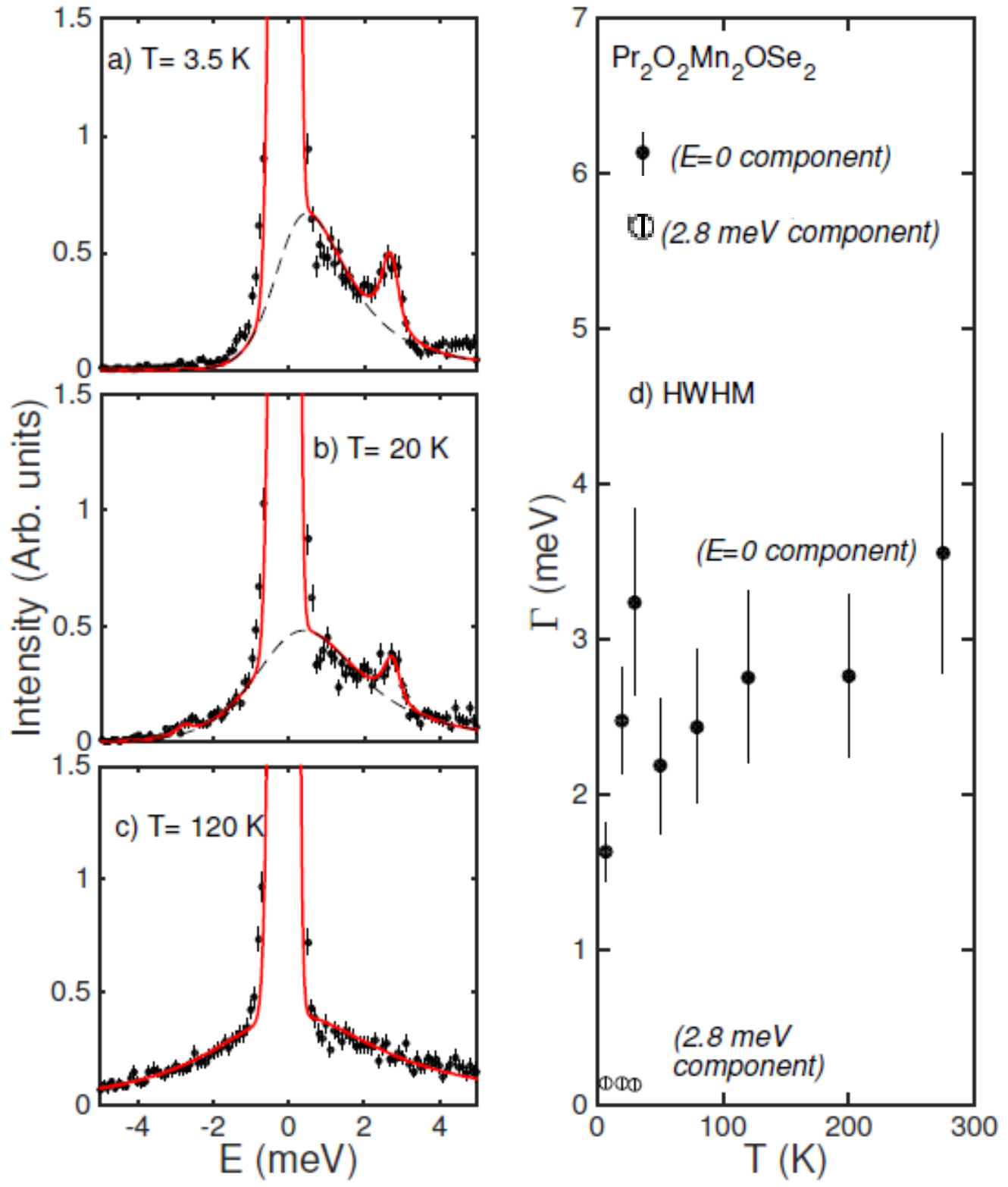

Figure 11 The lowest energy CEF in $Pr_2O_2Mn_2OSe_2$: the linewidth of the excitation is shown as a function of temperature (panel (d)).

**Discussion**

From INS experiments, we have observed significant temporal broadening of the lowest energy $Pr^{3+}$ ligand field excitations at all temperatures in $Pr_2O_2Mn_2OSe_2$, and only at temperatures above ~30 – 50 K in $Pr_2O_2Fe_2OSe_2$. Strong dampening of crystal field excitations has been observed previously in several systems including $PrOs_4Sb_{12}$[64] and Gd-based metals[65] and theories for this broadening and its temperature dependence have been developed in the context of $PrTl_3$.[66] However, all these theories involve a coupling to the electronic degrees of freedom and are difficult to apply to the $Pr_2O_2M_2OSe_2$ oxyselenides which are semimetallic/insulating with energy gaps ~1 eV. Any coupling between crystal electric field excitations and electronic degrees of freedom in $Pr_2O_2M_2OSe_2$ materials are unlikely to be on the meV energy scale probed in our experiments.

As discussed above, NPD experiments indicate that $Pr_2O_2M_2OSe_2$ ($M$ = Mn, Fe) undergo distortions from tetragonal to orthorhombic symmetry at low temperatures driven by the non-Kramers $Pr^{3+}$ ion, with the distortion of the $Fe^{2+}$ analogue much smaller than for the $Mn^{2+}$ phase. Such an orthorhombic distortion would split the lowest energy excited crystal field doublet, which may manifest as a broadening of the lowest energy crystal field excitation. However, no such broadening or splitting is observed in the low temperature data for $Pr_2O_2Fe_2OSe_2$ and broadening is only observed in its higher temperature tetragonal phase. Significant broadening is observed for $Pr_2O_2Mn_2OSe_2$ at all temperatures with an overdamped component at the lowest temperatures studied. It is therefore difficult to explain the experimental measurements of the crystal fields for these materials in terms of their structural behavior.

High temperature splitting of degenerate crystal field excitations has been observed for $CeAl_2$[67] and was attributed to a dynamic second order Jahn-Teller effect. Whilst it may be tempting to attribute the high temperature broadening for $Pr_2O_2Fe_2OSe_2$ to such effects, it is not consistent with the $Pr_2O_2Mn_2OSe_2$ data, for which broadening was observed at all temperatures. Similar high temperature splitting has also been observed for PrFeAsO,[46] and fluorine-doped $PrFeAsO_{1-x}F_x$ showed broadened and split crystal field excitations due to electrical inhomogeneity or possible local electronic disorder.[68] However, this explanation is also inconsistent with the $Pr_2O_2M_2OSe_2$ properties discussed in this work which have a high degree of structural order and are stoichiometric or very close to stoichiometric (from high resolution NPD experiments described here and in reference [14]).

Another possible mechanism for the observed broadening is coupling between ligand field excitations and magnon excitations on the transition metal sites, as observed for $Tb_3Fe_5O_{12}$.[69] Magnon excitations could provide decay channels analogous to electronic excitations,[66] given constraints due to momentum and energy transfer. These conditions are met in $Pr_2O_2Mn_2OSe_2$ at all temperatures, as $Mn^{2+}$ magnon excitations extend to low energy transfers (limited by the resolution) and result in spectral weight as $Q \rightarrow 0$ due to the competing exchange interactions. These magnon excitations could therefore provide decay routes for the $Pr^{3+}$ crystal field at all temperatures. A similar mechanism could explain the broadening of the $Pr^{3+}$ excitations only above 30 – 50 K in $Pr_2O_2Fe_2OSe_2$: gapped magnetic excitations with $\Delta \approx 5$ meV ≈ 58 K mean that such decay channels are only thermally and kinetically accessible at these high temperatures. Our results apply to $Pr^{3+}$ oxyselenides but it's also interesting to consider other lanthanide oxyselenides: $^{139}La$ spin relaxation in $La_2O_2Fe_2OSe_2$ has been investigated by NMR[12] and an anomaly in the relaxation rate was observed at ~50 K, the same temperature below which ligand field excitations sharpen in $Pr_2O_2Fe_2OSe_2$. The temperature dependence of the NMR relaxation rate was analysed in terms of thermally activated gapped excitations, consistent with the interpretation involving coupled crystalline and transition metal magnetic excitations proposed here.

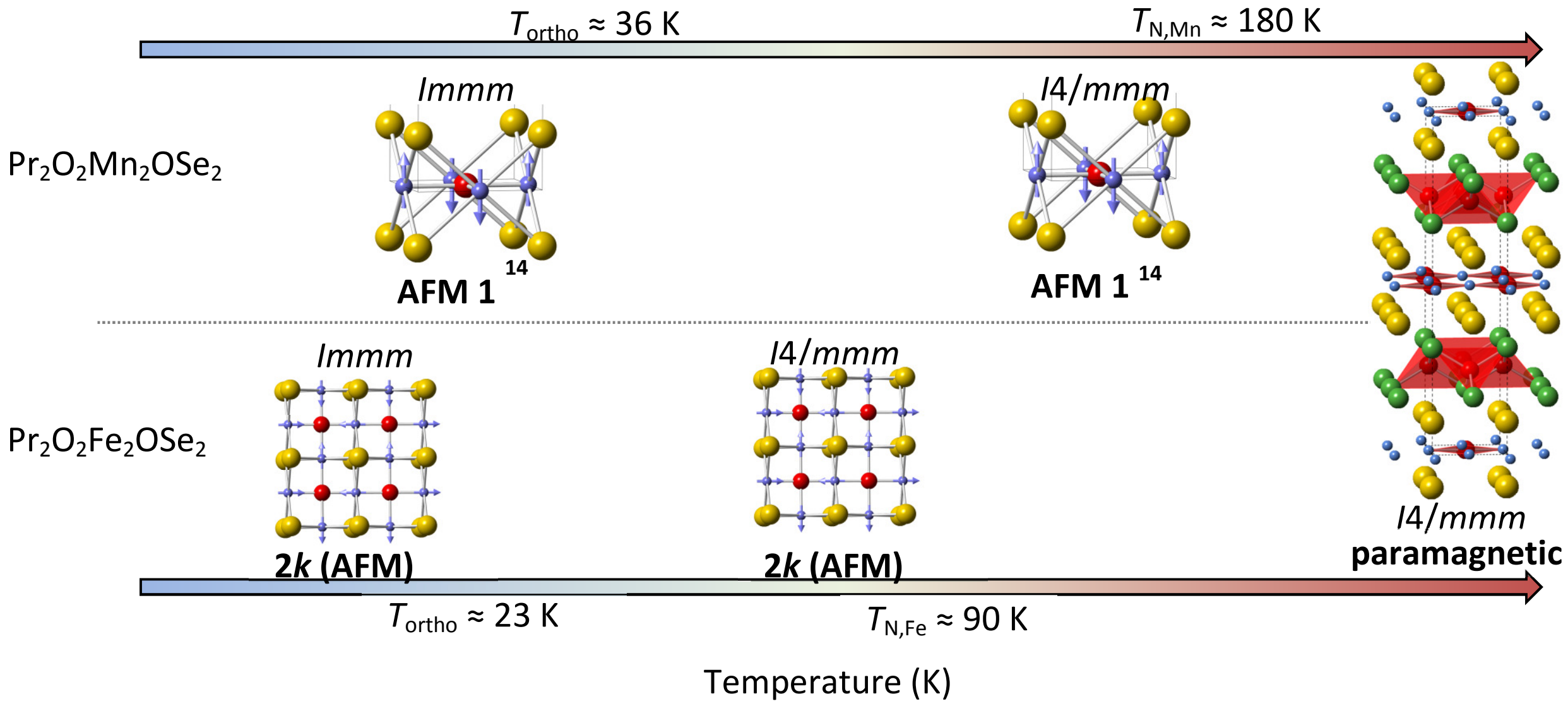

Figure 12 Schematic illustrating phase transitions of $Pr_2O_2M_2OSe_2$ (in zero-field) as a function of temperature (temperature axis not to scale), showing the tetragonal paramagnetic phase at high temperatures (right), cooling through $T_N$ to the tetragonal AFM phases (AFM1 for $M$ = Mn,[14] $2k$ for $M$ = Fe), and structural distortion to the orthorhombic *Immm* phase at low temperatures (left). We note that this orthorhombic distortion is much more subtle for $M$ = Fe than for $M$ = Mn and that in zero-field, we have no evidence for long-range ordering of $Pr^{3+}$ moments below this structural transition temperature.

## Conclusion

The crystal and magnetic structures of $Pr_2O_2M_2OSe_2$ ($M$ = Mn, Fe) have been explored using neutron scattering experiments in zero field (summarised in Figure 12) and applied magnetic field ($M$ = Fe). Whilst the magnetic structures and exchange interactions are similar to those observed for other lanthanide analogues,[10, 14-15, 17, 36-37] the structural chemistry is influenced by the symmetry-lowering structural distortion driven by the non-Kramers $Pr^{3+}$ ion. Analysis of crystal field excitations suggests some coupling between $Pr^{3+}$ crystal field excitations and transition metal magnetic excitations. This may give some insight into the non-Heisenberg coupling between the lanthanide and transition metal sublattices reported for oxypnictides and oxychalcogenides.[12, 38] $Pr_2O_2Fe_2OSe_2$ is more resistant to the $Pr^{3+}$-driven orthorhombic distortion than the $M$ = Mn analogue,[14] and this resistance is likely to be due to its magnetic behaviour. Further INS experiments to study CEFs with temperature in other magnetic systems would add to our understanding of this coupling.

## Acknowledgements

We're grateful to The Carnegie Trust, EPSRC (EP/J011533/1), STFC and University of Kent for funding. We thank the ISIS neutron source for the beamtime and Dr Pascal Manuel and Dr Laurent Chapon for assistance collecting NPD data.

**Supplementary Materials**

SM1: (a), (b), (c) Unit cell parameters of $Pr_2O_2Fe_2OSe_2$ on cooling in zero applied magnetic field from sequential Rietveld refinements using high resolution NPD data fitted with a tetragonal unit cell, (d) d-spacing dependent peak shape term as a function of temperature from sequential refinements using *I*4/*mmm* Pawley phase with values for *h*00/0*k*0, *hhl* and other reflections shown in red, green and blue, respectively; (e) *z* factional coordinate for O(2) site from sequential Rietveld refinements for *Immm* model.

a)

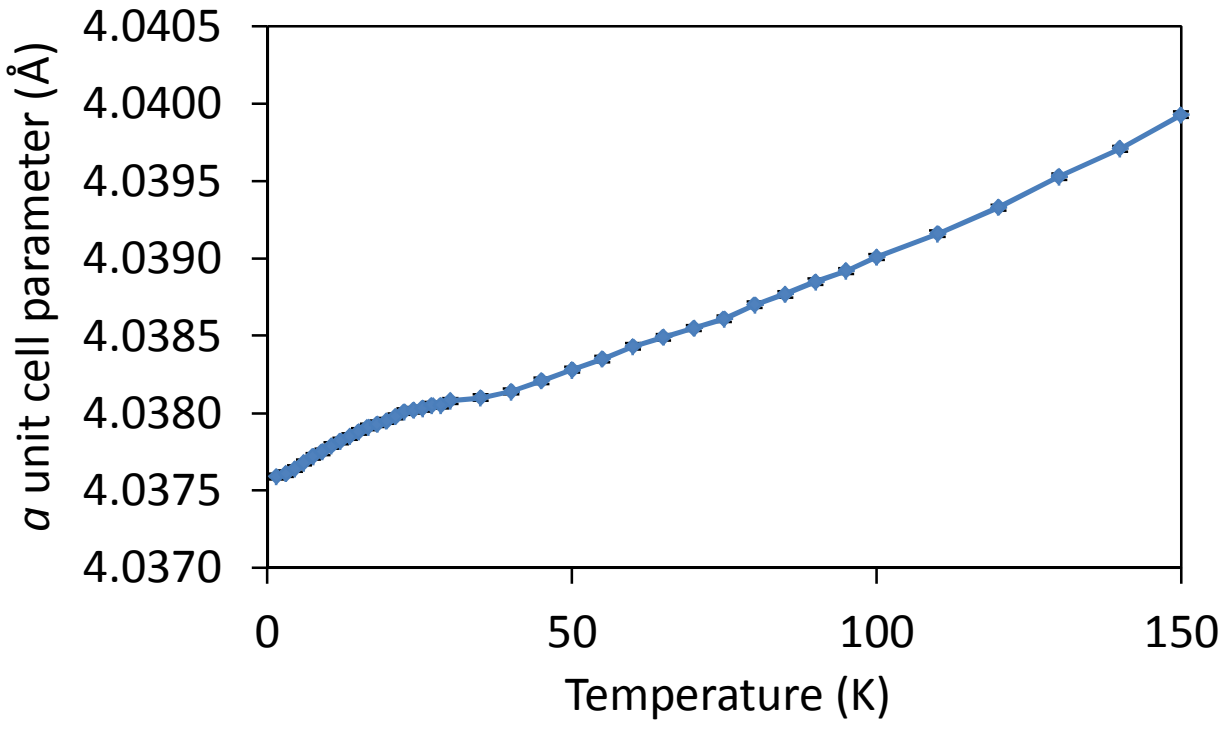

b)

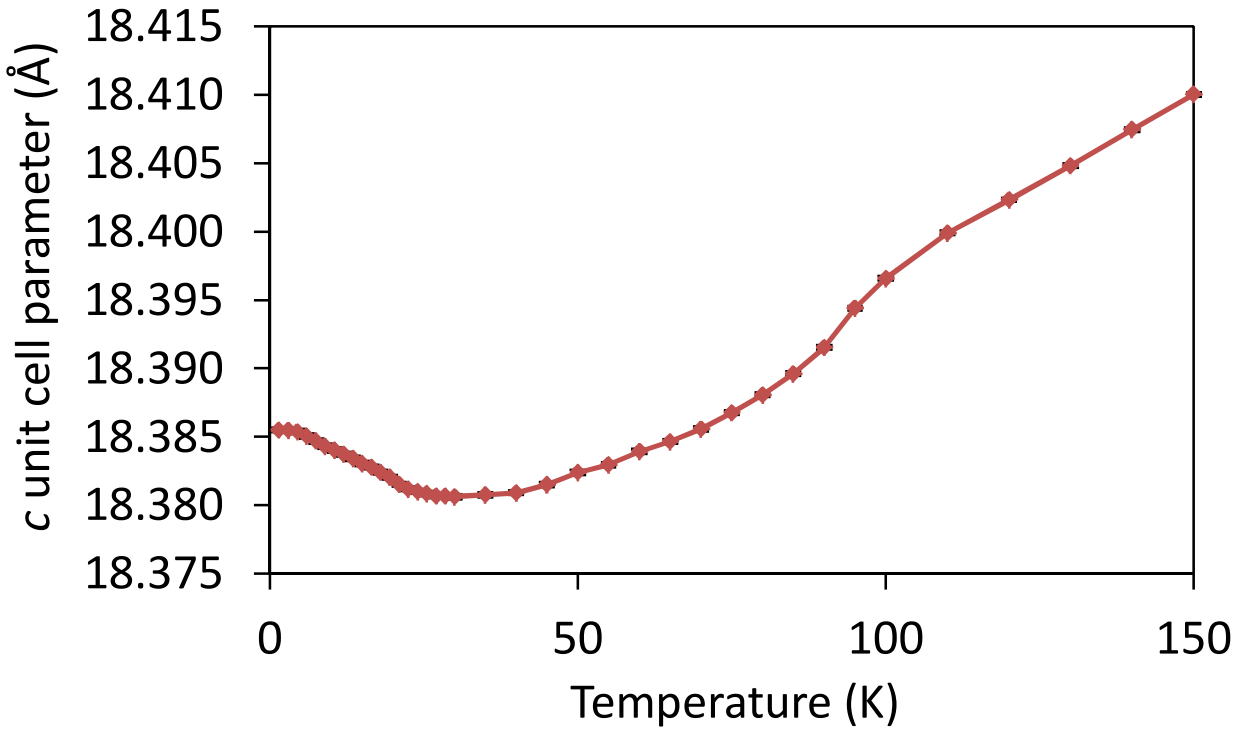

c)

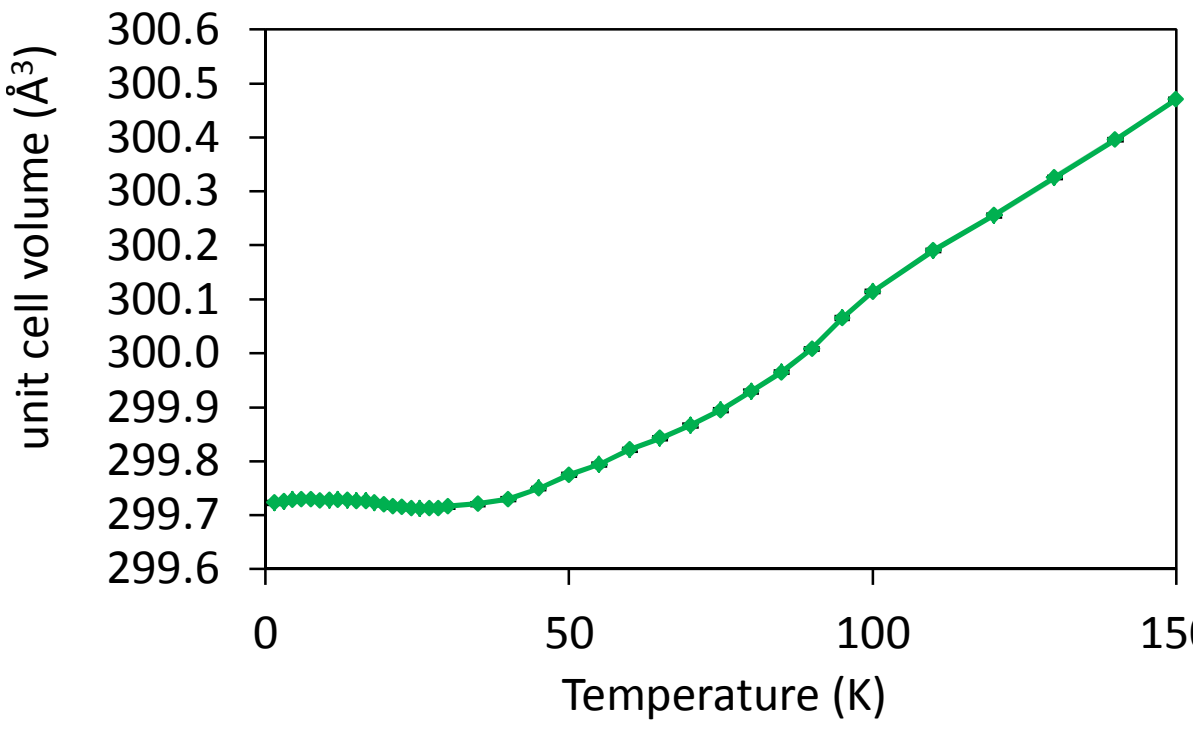

d)

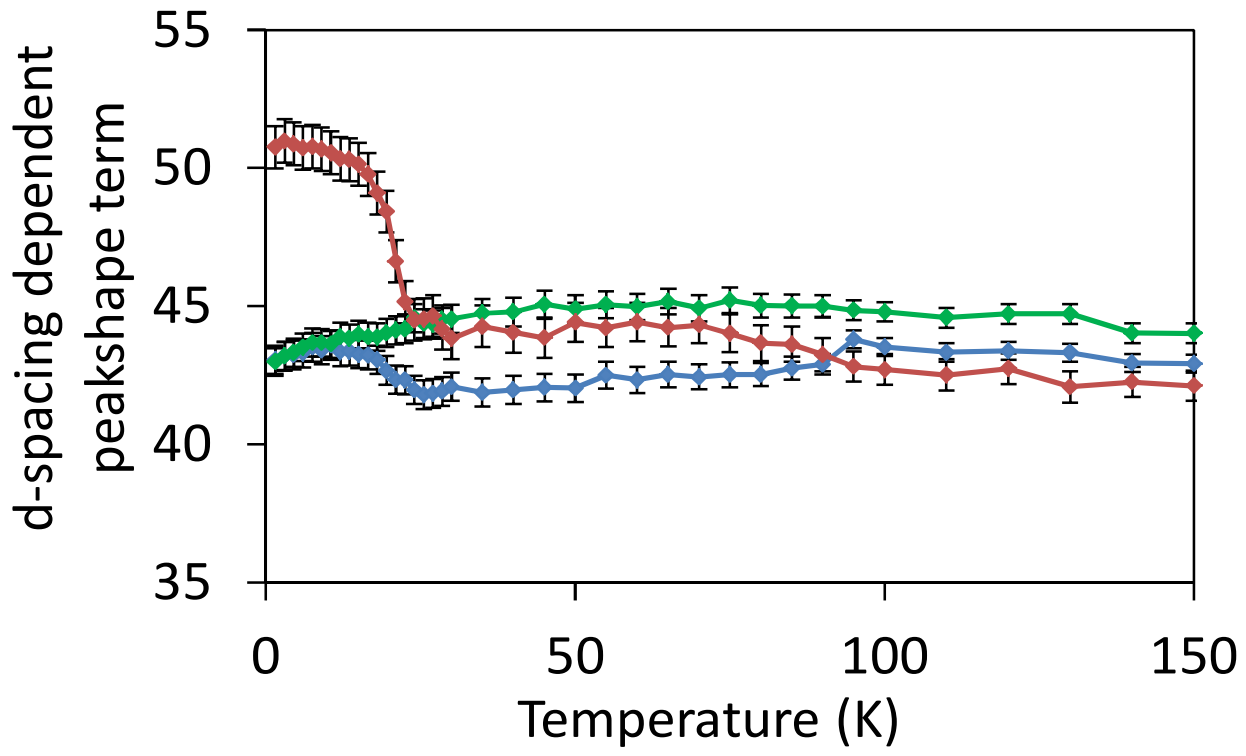

e)

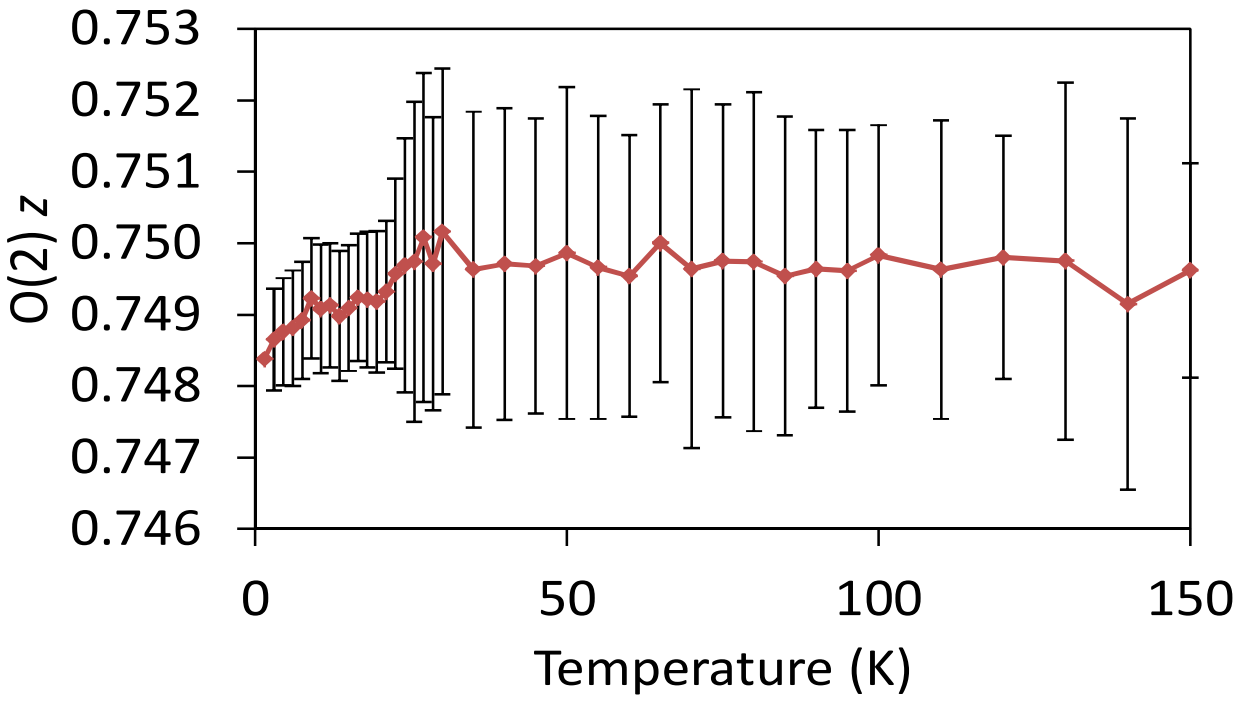

SM2: Rietveld refinement details for $Pr_2O_2Fe_2OSe_2$ using tetragonal *I*4/*mmm* model to fit NPD data collected at 2 K in the absence of an applied magnetic field.

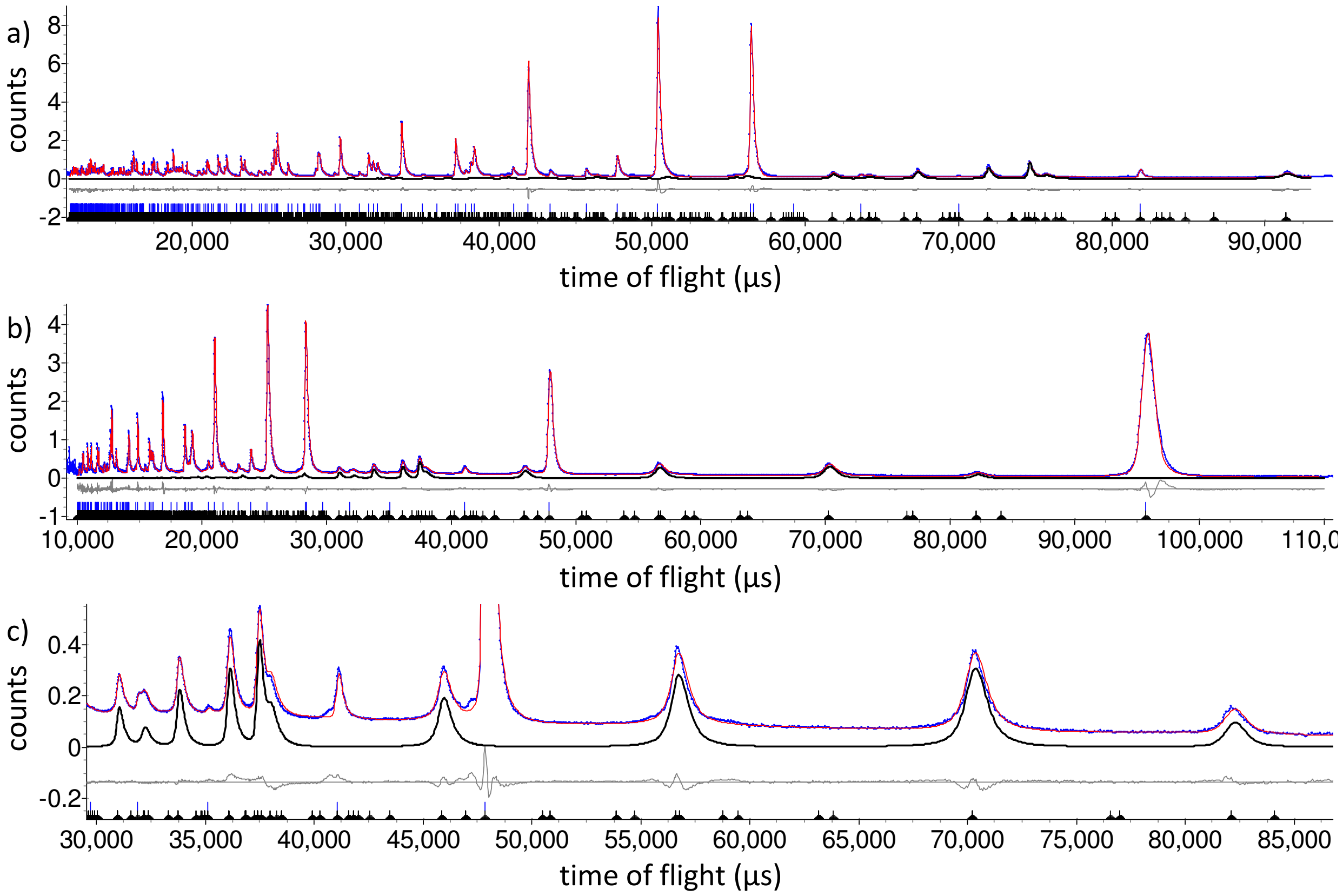

Figure SM2.1 (Color online) Rietveld refinement profiles from combined refinement (a) using 153° bank data and (b) 59° bank data collected for $Pr_2O_2Fe_2OSe_2$ at 1.5 K; (c) shows the higher *d*-spacing region of 59° bank data highlighting the magnetic reflections. Observed and calculated (upper) and difference profiles are shown by blue points, and red and gray lines, respectively. Magnetic intensity is highlighted by solid black line.

Table SM2.1 Details from Rietveld refinement using NPD data collected at 1.5 K for $Pr_2O_2Fe_2OSe_2$. The refinement was carried out with the nuclear structure described by space group *I*4/*mmm* with *a* = 4.0366(1) Å and *c* = 18.3812(9) Å. The magnetic scattering was fitted by a second magnetic-only phase with *a*, *b*, and *c* unit cell parameters constrained to be twice those of the nuclear phase; $R_{wp}$ = 5.128%, $R_p$ = 4.865%, and $\chi^2$ = 11.04.

| **Atom** | **Site** | ***x*** | ***y*** | ***z*** | $U_{iso}$ **× 100 (Å$^2$)** | **Moment ($\mu_B$)** |
|---|---|---|---|---|---|---|
| Pr | 4*e* | 0.5 | 0.5 | 0.18608(6) | 0.29(4) | |
| Fe | 4*c* | 0.5 | 0 | 0 | 0.21(1) | 3.365(9) (*xy*) |
| Se | 4*e* | 0 | 0 | 0.09807(4) | 0.00(2) | |
| O(1) | 4*d* | 0.5 | 0 | 0.25 | 0.30(2) | |
| O(2) | 2*b* | 0.5 | 0.5 | 0 | 0.50(4) | |

SM3: Unit cell parameters of $Pr_2O_2Fe_2OSe_2$ from sequential Rietveld refinements using high resolution NPD data: (a) and (b) in 5 T applied field on cooling; (c) and (d) at 2 K as a function of applied field.

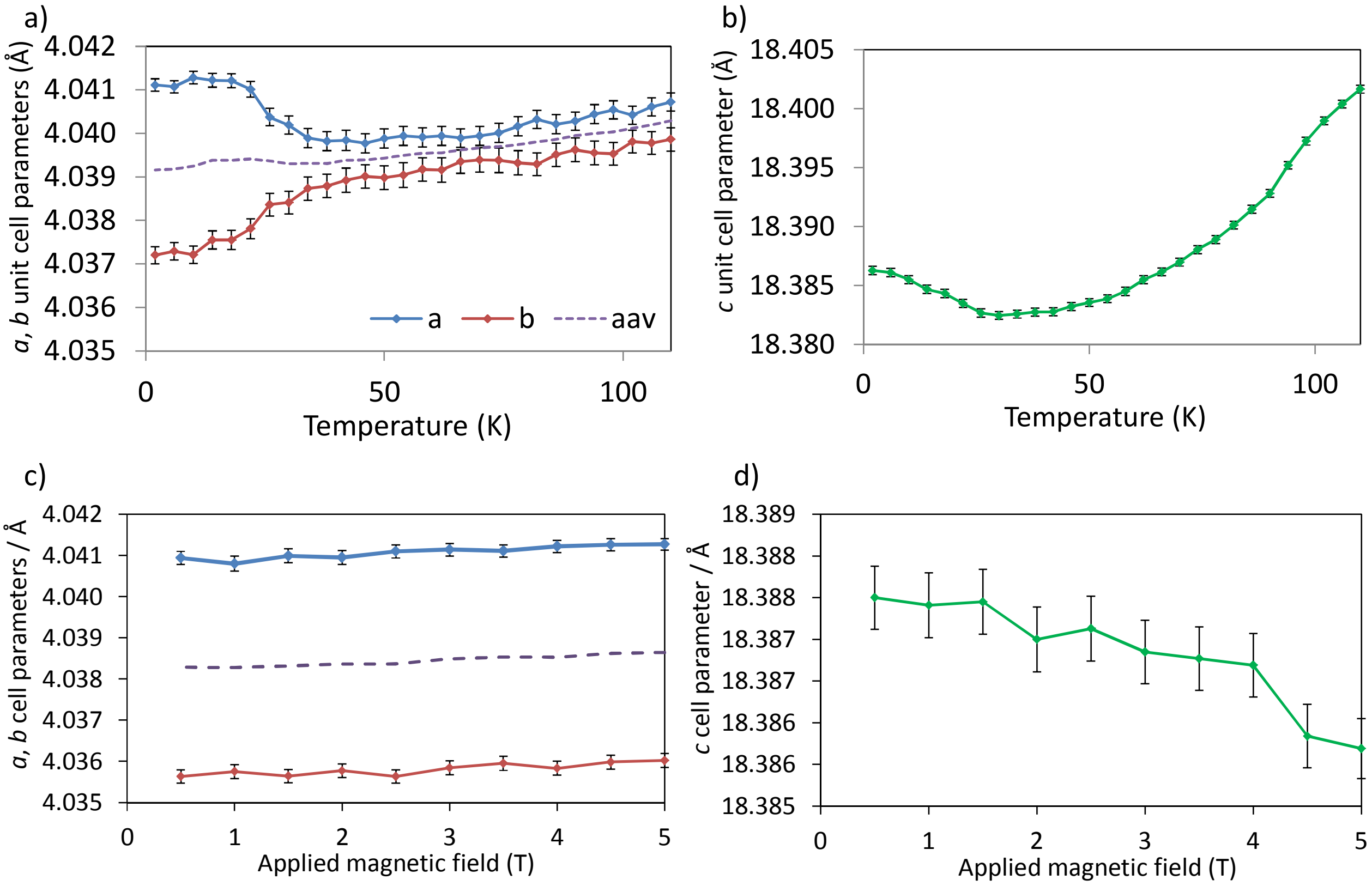

SM4: A weak Warren-like peak is observed for $Pr_2O_2Fe_2OSe_2$ for only a few Kelvin above $T_N$ in zero applied magnetic field (a). The figure shows this most clearly in data collected at 95 K (green) in the top panel. No such peak is observed in data collected in 5 T applied magnetic field (b) and weak magnetic Bragg peaks are observed in 94 K (green) data in the bottom panel. The main magnetic Bragg reflections for the 2-*k* vector magnetic structure are labelled.

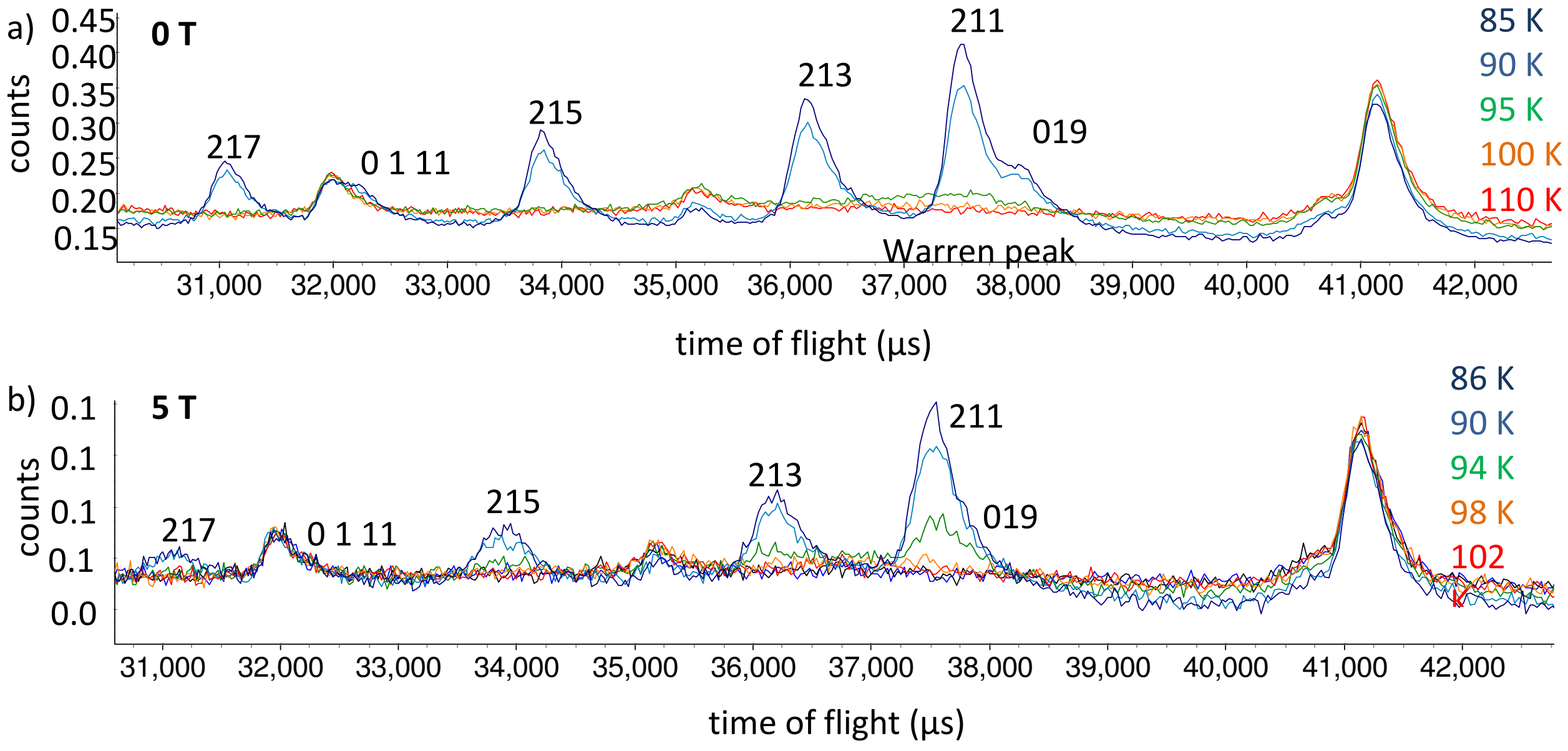

SM5: Magnetic property measurements on $Pr_2O_2Fe_2OSe_2$

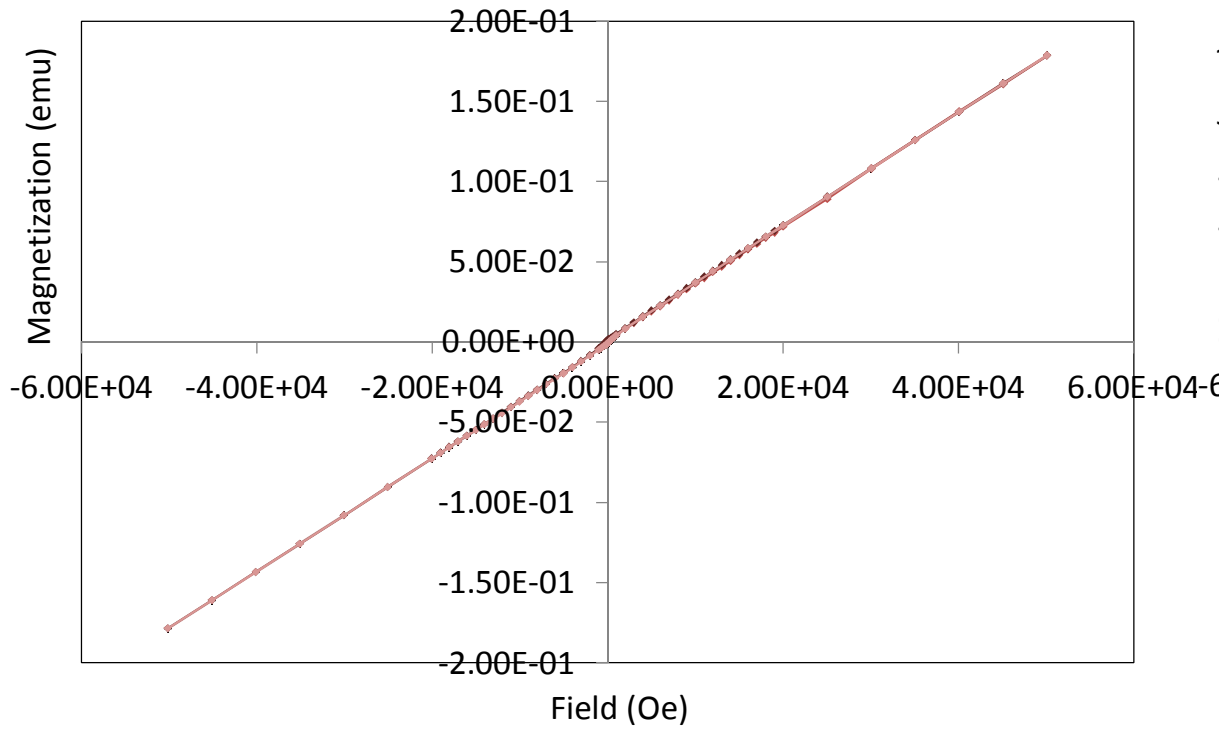

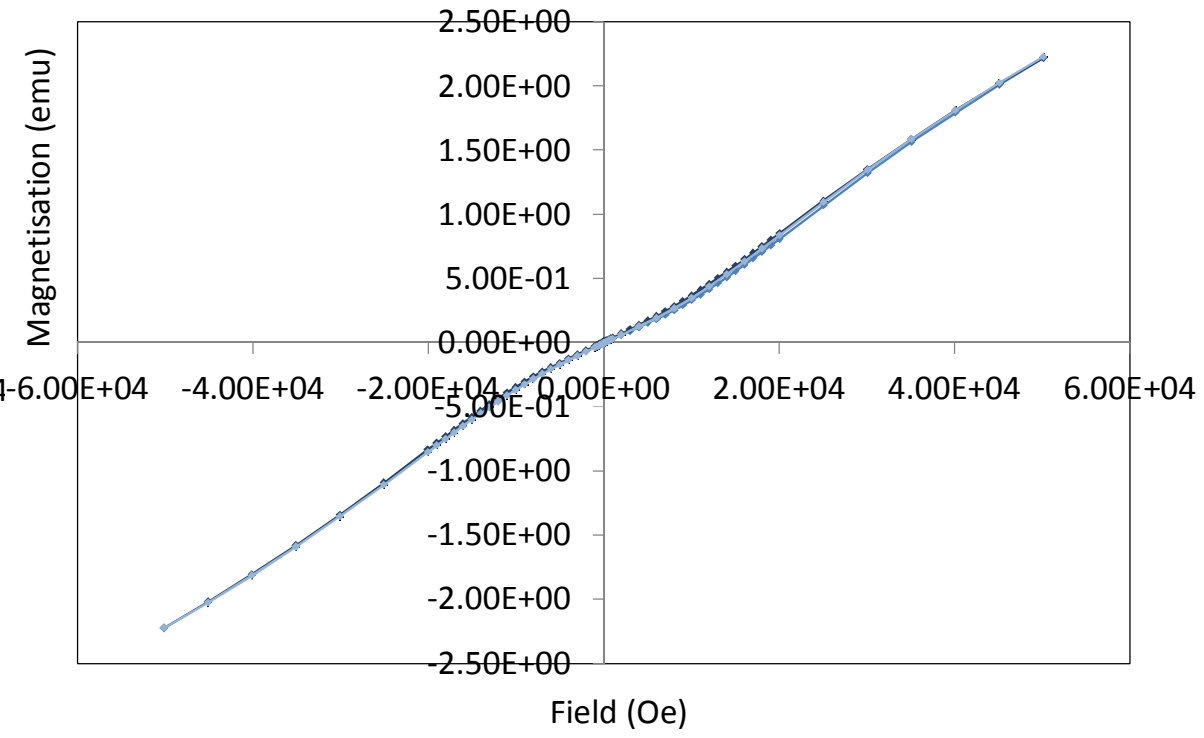

Field sweep measurements at 295 K were linear and passed through the origin, however, similar measurements on the same sample at 12 K showed some field dependence: although these data passed through the origin, the magnetisation was not linear with increasing field. This suggested the possibility of some metamagnetic behaviour at low temperatures and prompted us to collect susceptibility data at low and high fields, below.

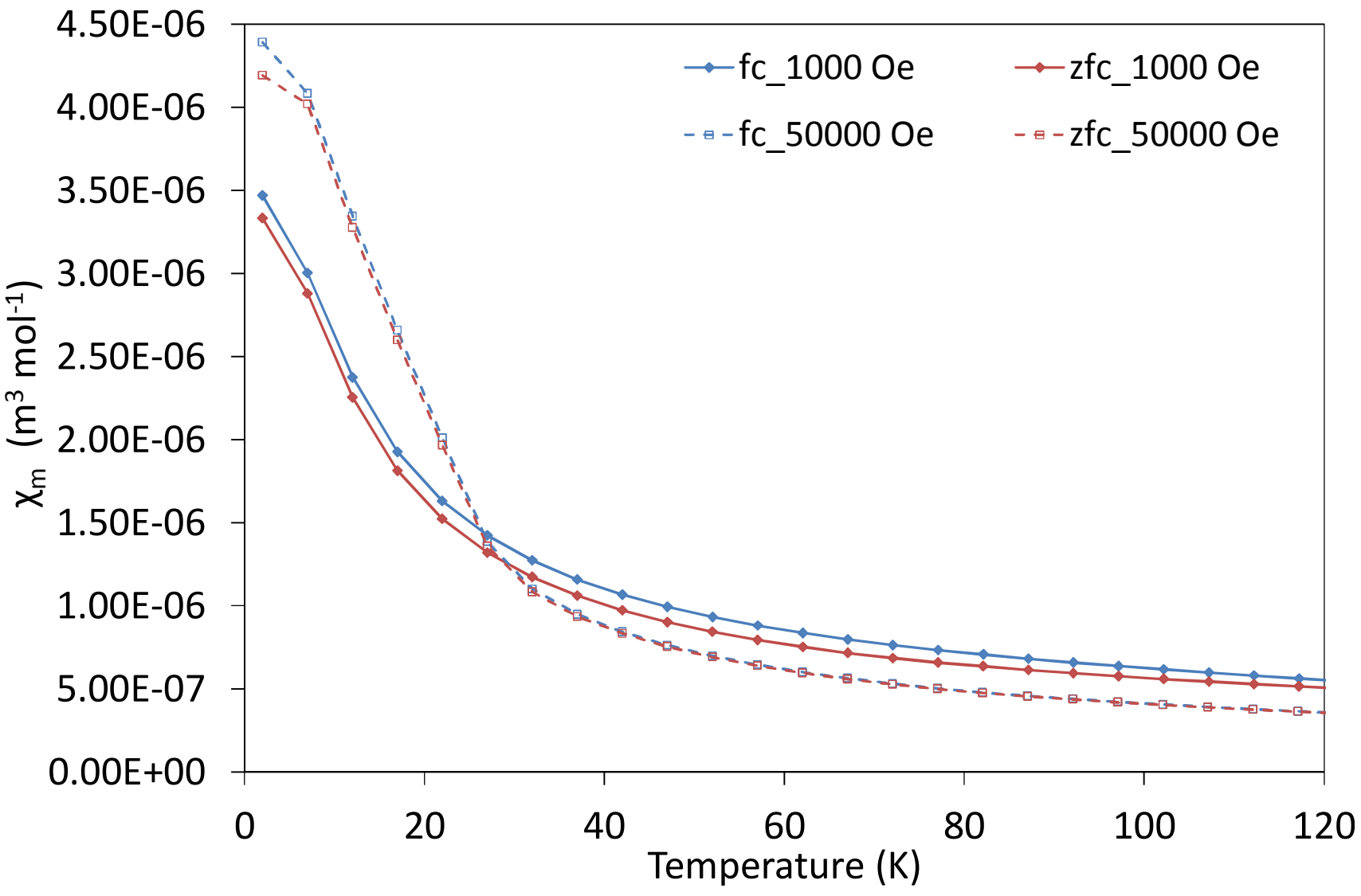

Magnetic susceptibility data were collected for $Pr_2O_2Fe_2OSe_2$ in applied magnetic fields of 1000 Oe (solid points) and 50000 Oe (open points) showing field dependence below ~22 K, consistent with reports of metamagnetism by Ni et al.[36]

SM6: Intensities of magnetic Bragg reflections on cooling in zero applied magnetic field from sequential refinements using a Pawley phase to fit peak intensities.

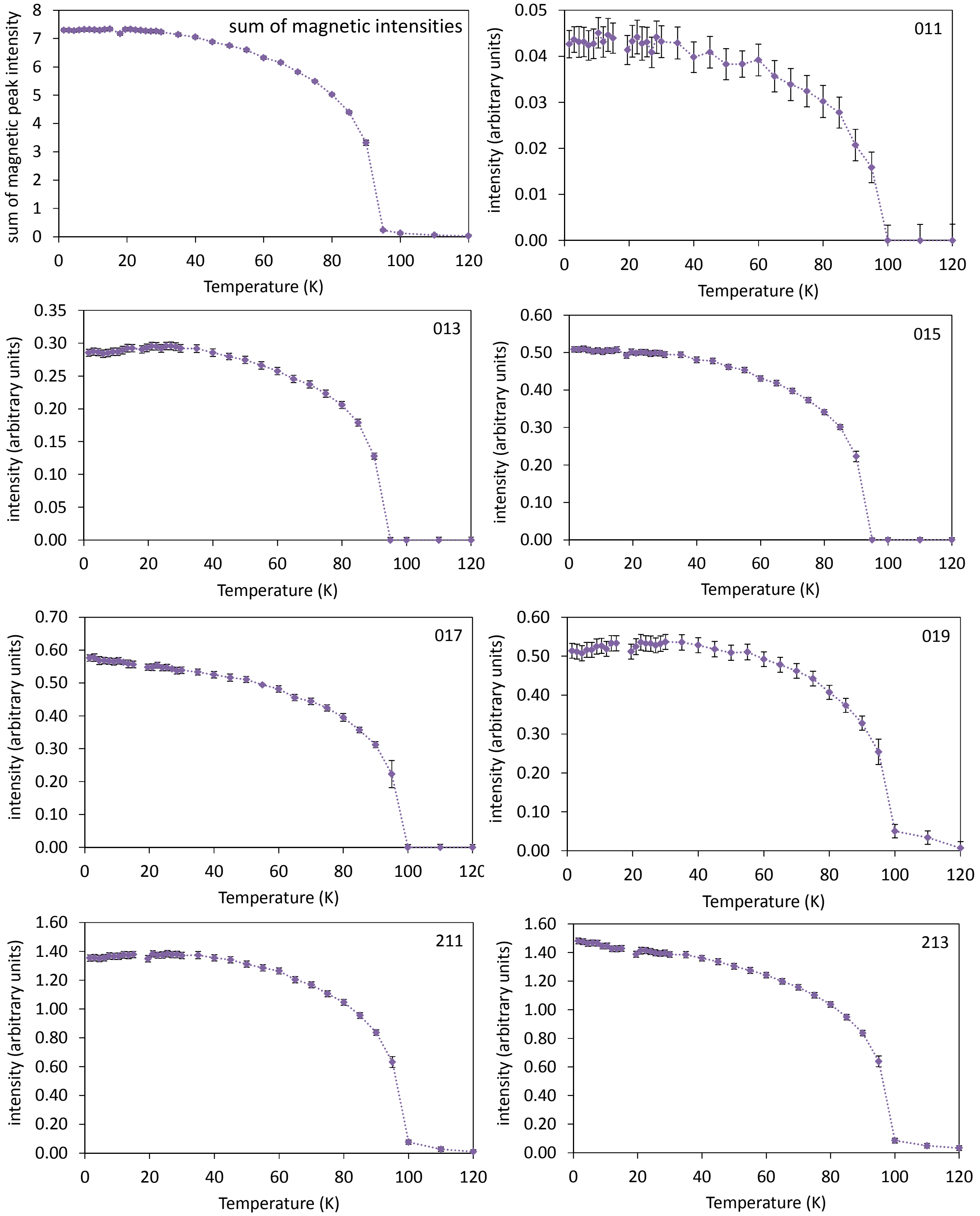

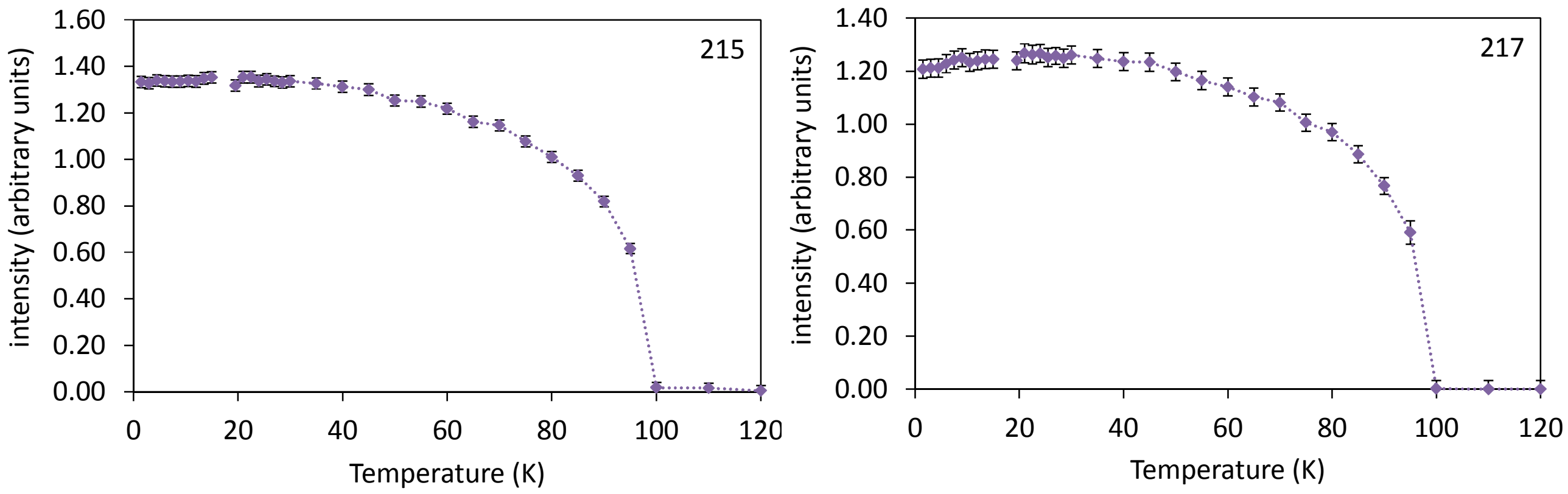

SM7: Fit to magnetic Bragg reflections at 1.5 K in zero field

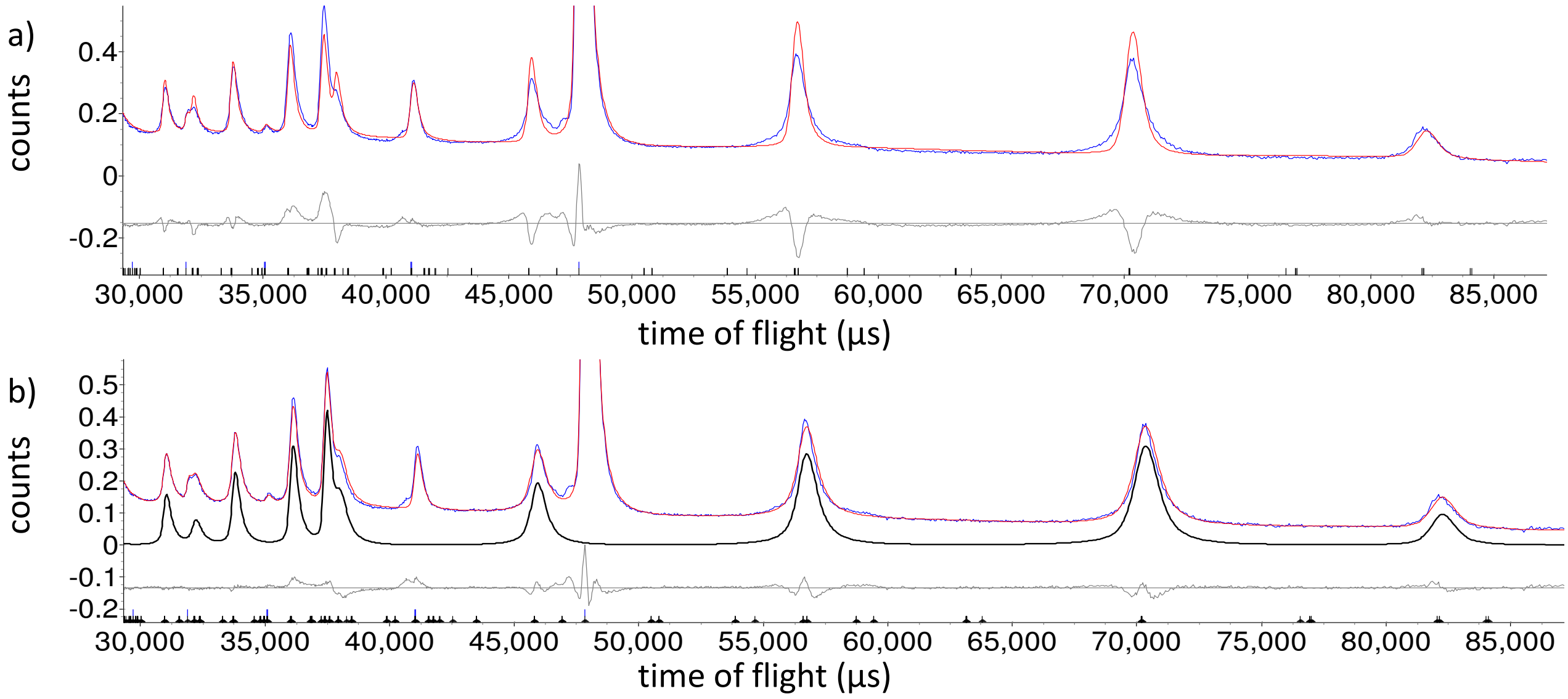

Figure SM7 (Color online) Rietveld refinement profiles from combined refinement using 153° bank and 59° bank data collected for $Pr_2O_2Fe_2OSe_2$ at 1.5 K showing the higher *d*-spacing region of 59° bank data highlighting the magnetic reflections with a) fitted with no anisotropic peak broadening and b) with anisotropic peak broadening fitted by an expression for antiphase boundaries perpendicular to *c*. Observed and calculated (upper) and difference profiles are shown by blue points, and red and grey lines, respectively. Magnetic intensity is highlighted by solid black line.

SM8: Intensities of magnetic Bragg reflections on cooling in 5 T applied field from sequential refinements using a Pawley phase to fit peak intensities. Plots show magnetic Bragg peaks observed for $T < 94$ K due to the 2-k magnetic ordering on the Fe sublattice (purple), and additional reflections that appear for $T <$ ~26 K (green).

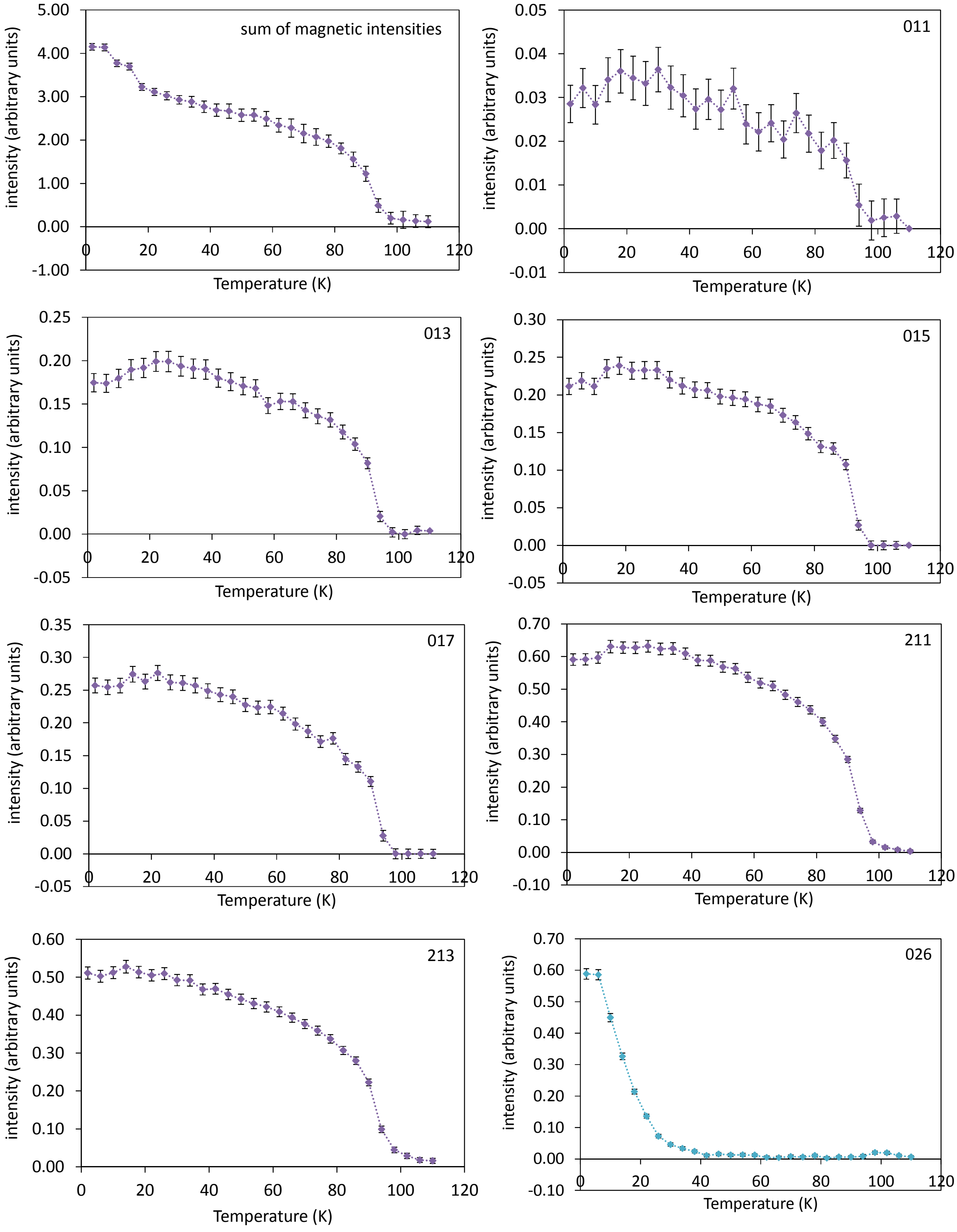

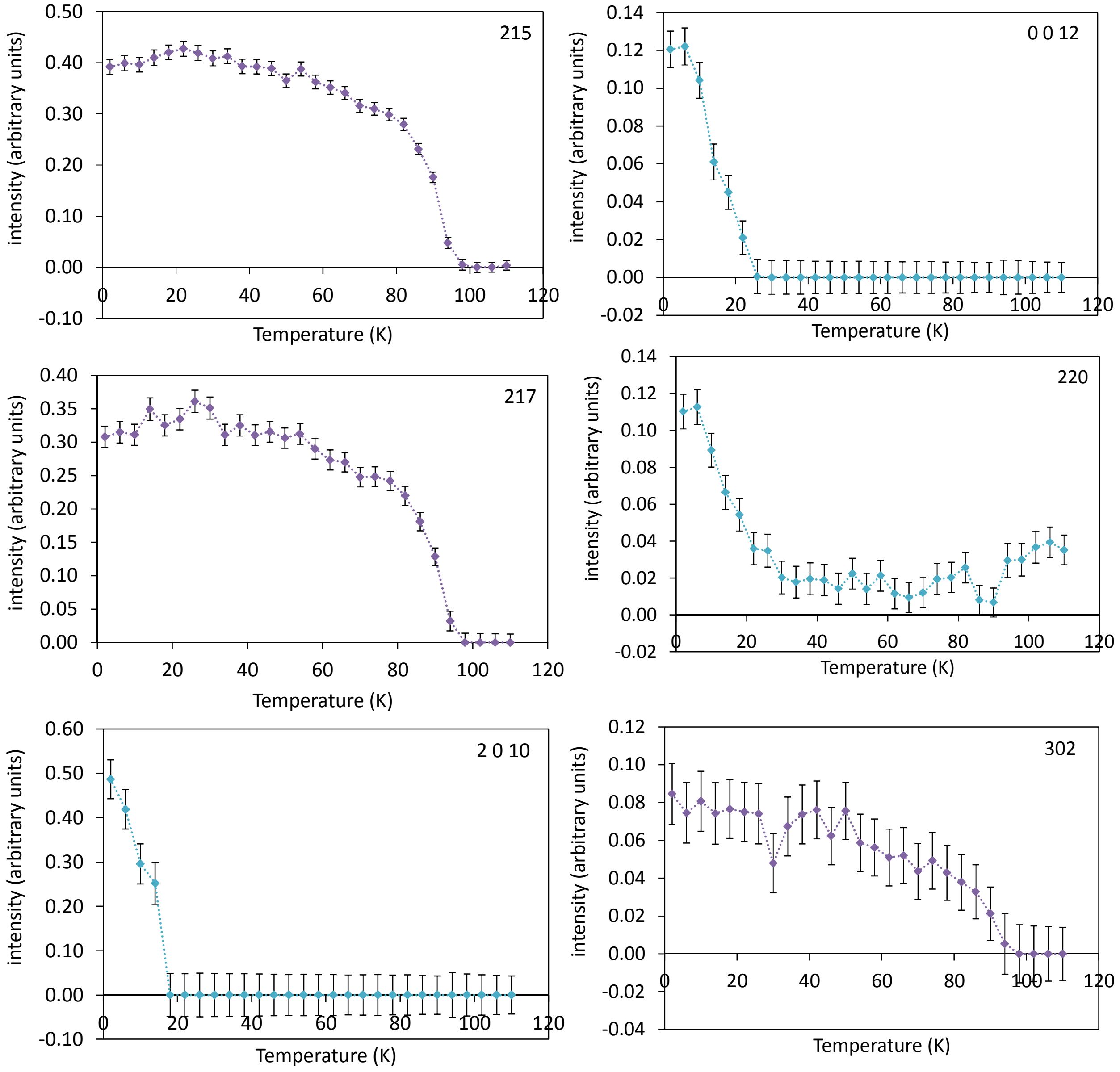

SM9: Magnetic structure of $Pr_2O_2Fe_2OSe_2$ from NPD refinement using data at 1.5 K in 5 T applied magnetic field.

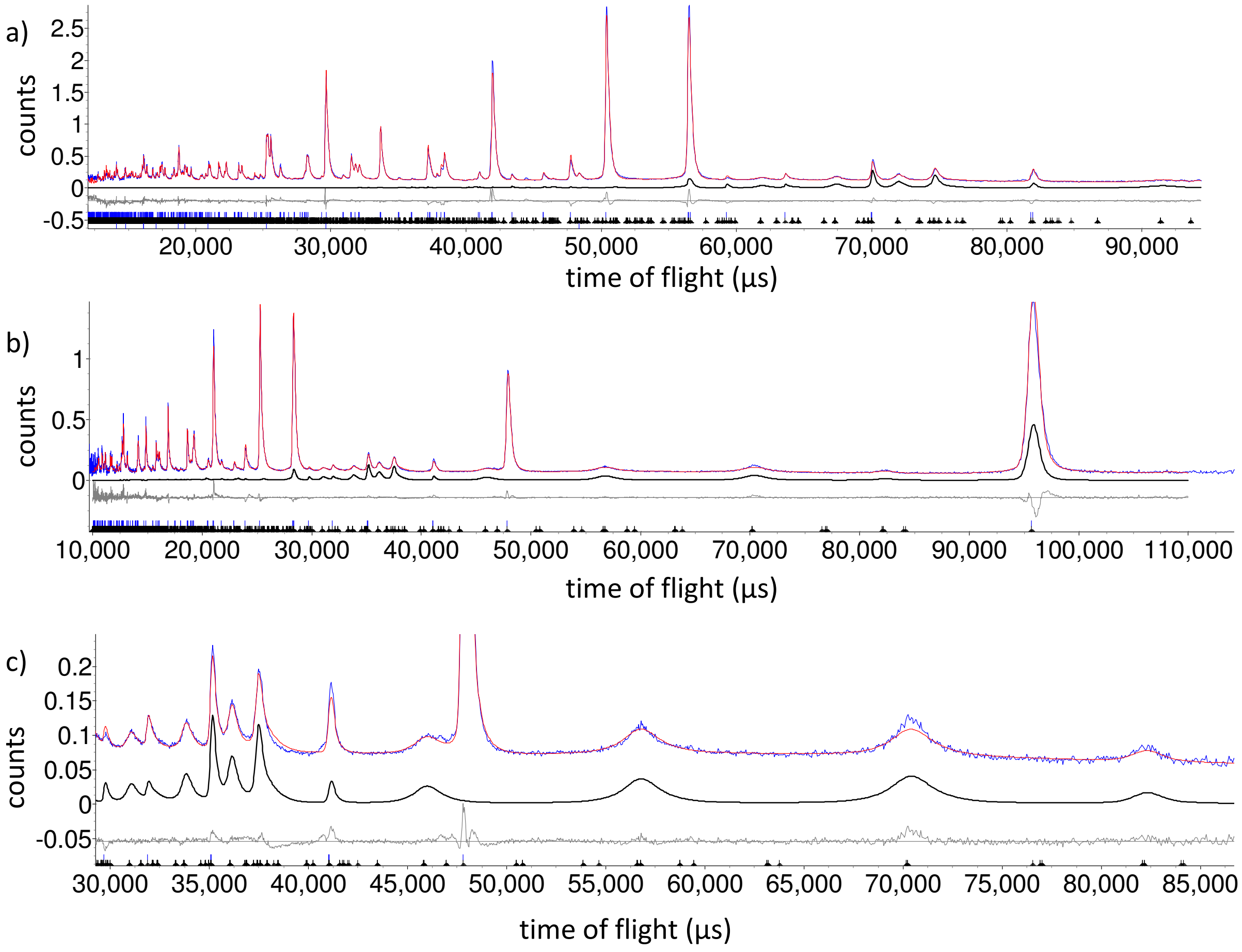

Figure SM9.1 (Color online) Rietveld refinement profiles from combined refinement (a) using 153° bank data and (b) 59° bank data collected for $Pr_2O_2Fe_2OSe_2$ at 2 K in 5 T applied magnetic field; (c) highlights the higher *d*-spacing region of 59° bank data emphasising the magnetic reflections. Observed and calculated (upper) and difference profiles are shown by blue points, and red and grey lines, respectively.

Table SM9 Details from Rietveld refinement using NPD data collected at 2 K for $Pr_2O_2Fe_2OSe_2$ in 5 T applied magnetic field. The refinement was carried out with the nuclear structure described by space group *Immm* with *a* = 4.0388(4) Å, *b* = 4.0328(4) Å and *c* = 18.370(1) Å. The magnetic scattering was fitted by a second magnetic-only phase with *a*, *b*, and *c* unit cell parameters constrained to be twice those of the nuclear phase; $R_{wp}$ = 5.66%, $R_p$ = 5.73%, and $\chi^2$ = 6.44.

| **Atom** | **Site** | ***x*** | ***y*** | ***z*** | **$U_{iso}$ × 100 (Å$^2$)** | **Moment ($\mu_B$)** |
|---|---|---|---|---|---|---|
| Pr | 4*i* | 0 | 0 | 0.6855(1) | 1.0(1) | 2.21(1) (*y*) |
| Fe(1) | 2*d* | 0.5 | 0 | 0.5 | 0.80(5)* | 3.36(2) (*x*/*y*) |
| Fe(2) | 2*b* | 0 | 0.5 | 0.5 | 0.80(5)* | 3.36(2) (*x*/*y*) |
| Se | 4*i* | 0 | 0 | 0.09896(7) | 0.22(5) | |
| O(1) | 4*j* | 0.5 | 0 | 0.7538(3) | 0.17(7) | |
| O(2) | 2*c* | 0.5 | 0.5 | 0 | 0.28(8) | |

* $U_{iso}$ for Fe(1) and Fe(2) constrained to be the same.

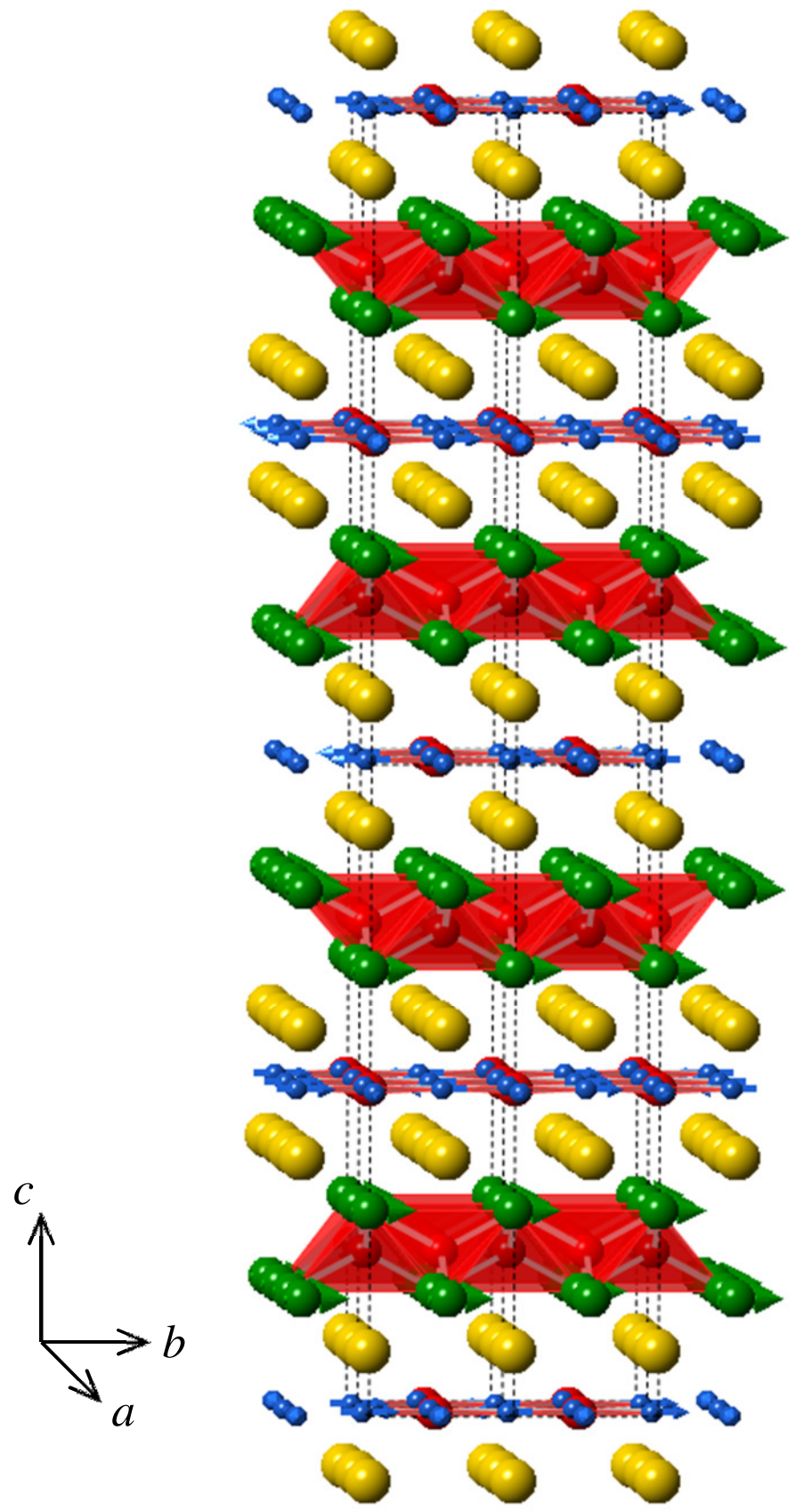

Figure SM9.2 (Color online) Illustration of the magnetic structure of $Pr_2O_2Fe_2OSe_2$ determined from refinement using neutron powder diffraction data at 1.5 K in 5 T applied magnetic field. Fe, Pr, O and Se sites are shown in blue, green, red and yellow with $Fe^{2+}$ and $Pr^{3+}$ moments shown by blue and green arrows, respectively.

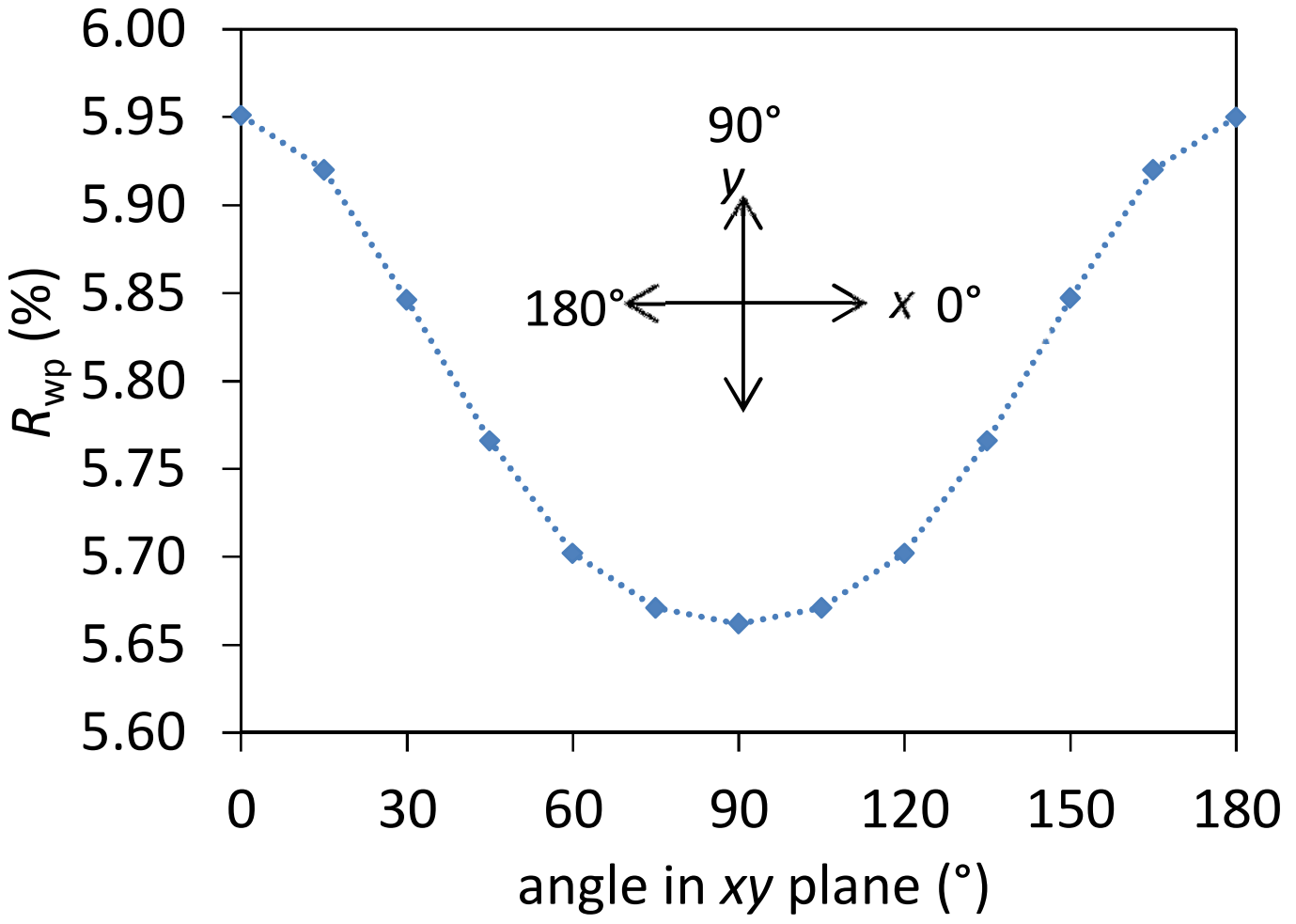

Figure SM9.3 (Color online) Plot shows variation in fit to observed NPD data (collected at 2 K in 5 T applied magnetic field) as a function of direction of $Pr^{3+}$ moments within *xy* plane (0° indicates moments along *x* direction, 90° indicates moments along *y*) showing that the best fit is obtained with moments oriented along the shorter *y* axis.